\documentclass[useAMS,usenatbib]{mnras}
\usepackage[utf8]{inputenc}
\usepackage[english]{babel}
\usepackage{graphicx}
\usepackage{caption}
\usepackage{hyperref}
\usepackage[section] {placeins}
\usepackage{amsmath, amssymb}
\usepackage{txfonts}
\usepackage{gensymb}
\usepackage{color}
\usepackage{multirow}
\usepackage{makeidx}
\usepackage{xspace}
\usepackage{float}
\usepackage{verbatim}
\usepackage[normalem]{ulem}
\usepackage{subcaption}
\usepackage{siunitx}
\usepackage{lmodern} 
\newcommand{\refsec}[1]{\S~\ref{#1}}
\newcommand{\reffig}[1]{Fig.~\ref{#1}}

\newcommand{\reftab}[1]{Table~\ref{#1}}
\newcommand{\refapp}[1]{Appendix~\ref{#1}}

\newcommand{\Hb}{$\mathrm{H\beta}$\xspace}

\newcommand{\Ha}{$\mathrm{H\alpha}$\xspace}

\newcommand{\SII}[1]{$\mathrm{[S\,II]\lambda#1}$\xspace}

\newcommand{\nii}{\rm{[\ion{N}{II}]$\lambda$6584}}
\newcommand{\ha}{\rm H{$\alpha$}}

\newcommand{\siii}{\rm{[\ion{S}{II}]$\lambda\lambda$6717,6731}}

\def\Mgb{$\rm{Mg}b$\xspace}
\def\Mgbx{$\rm{Mg}b(\mathbf{x})$\xspace}
\def\SFR{$\mathrm{SFR}$\xspace}
\def\sSFR{$\mathrm{sSFR}$\xspace}

\def\ZH{$\rm{[Z/H]}$\xspace}

\def\re{$\mathrm{R_e}$\xspace}
\def\logre{$\log \, \mathrm{R_e}$\xspace}

\def\M{$\rm{M}$\xspace}

\def\Pot{$\Phi$\xspace}
\def\Sig{$\Sigma$\xspace}
\def\Z{$\mathrm{O/H}$\xspace}
\def\Zx{$\mathrm{O/H}(\mathbf{x})$\xspace}
\def\MZR{$\mathrm{M}$-$\mathrm{O/H}$\xspace}
\def\PZR{$\mathrm{\Phi}$-$\mathrm{O/H}$\xspace}
\def\SZR{$\mathrm{\Sigma}$-$\mathrm{O/H}$\xspace}
\def\RZR{$\mathrm{R}$-$\mathrm{O/H}$\xspace}
\def\MstarRe{$\rm{M_*/R_e}$\xspace}
\def\MstarReT{$\rm{M_*/R_e^2}$\xspace}
\def\Mstar{$\rm{M_*}$\xspace}
\def\Sigstar{$\Sigma_*$\xspace}
\def\Sigstarx{$\Sigma_*(\mathbf{x})$\xspace}

\def\OH{$\mathrm{O/H}$\xspace}
\def\OHx{$\mathrm{(O/H)}(\mathbf{x})$\xspace}
\def\OHO3N2{$\mathrm{(O/H)_{O3N2}}$\xspace}
\def\OHN2{$\mathrm{(O/H)_{N2}}$\xspace}
\def\OHT04{$\mathrm{(O/H)_{T04}}$\xspace}

\def\Siggas{$\Sigma_\mathrm{gas}$\xspace}

\def\fgasx{$f_\mathrm{gas}(\mathbf{x})$\xspace}

\def\vesc{$\mathrm{V_{esc}}$\xspace}
\def\vescx{$\mathrm{V_{esc}(\mathbf{x})}$\xspace}

\def\rhosp{$\rho$\xspace}
\def\rhospred{$\rho_r$\xspace}

\def\logsig3{$\log \, \Sigma_3$\xspace}

\defcitealias{barone+2018}{B18}
\defcitealias{barrera-ballesteros+2018}{BB18}
\defcitealias{tremonti+2004}{T04}
\defcitealias{dopita+2016}{D16}
\defcitealias{sanchez+2017}{S17}

\title[The \PZR relation with aperture-matched sampling]{The gas-phase metallicities of star-forming galaxies in aperture-matched SDSS samples follow potential rather than mass or average surface density}

\author[Francesco~D'Eugenio~et al.]{\parbox{\textwidth}{
Francesco~D'Eugenio,$^{1,2,3}$\thanks{E-mail: francesco.deugenio@ugent.be}
Matthew~Colless,$^{1,2,4}$
Brent~Groves,$^{1,4}$
Fuyan~Bian$^{1}$\thanks{Stromlo Fellow}
and {Tania~M.~Barone}$^{1,2,4,5}$}
\vspace{0.4cm}
\\
\parbox{\textwidth}{
$^{1}$Research School of Astronomy and Astrophysics, Australian National University, Canberra, ACT 2611, Australia\\
$^{2}$Australian Research Council Centre of Excellence for All-sky Astrophysics (CAASTRO)\\
$^{3}$Sterrenkundig Observatorium, Universiteit Gent, Krijgslaan 281 S9, B-9000 Gent, Belgium\\
$^{4}$ARC Centre of Excellence for All Sky Astrophysics in 3 Dimensions (ASTRO 3D), Australia\\
$^{5}$Sydney Institute for Astronomy, School of Physics, The University of Sydney, NSW, 2006, Australia\\
}
}

\begin{document}

    \date{\today}

    \pagerange{\pageref{firstpage}--\pageref{lastpage}} \pubyear{1581}

    \maketitle

    \label{firstpage}

  \begin{abstract}

    We present a comparative study of the relation between the aperture-based
    gas-phase metallicity and three structural parameters of star-forming galaxies:
    mass ($\mathrm{M \equiv M_*}$), average potential ($\Phi \equiv \mathrm{M_*/R_e}$)
    and average surface mass density ($\Sigma \equiv \mathrm{M_*/R_e^2}$; where
    \re is the effective radius). We use a
    volume-limited sample drawn from the publicly available SDSS DR7, and base
    our analysis on aperture-matched sampling by selecting sets of galaxies
    where the SDSS fibre probes a fixed fraction of \re.
    We find that between 0.5 and 1.5~\re, the gas-phase metallicity correlates
    more tightly with \Pot than with either \M or \Sig, in that for all
    aperture-matched
    samples, the potential-metallicity relation has (i) less scatter, (ii) higher
    Spearman rank correlation coefficient and (iii) less residual trend with \re
    than either the mass-metallicity relation and the average surface
    density-metallicity relation.
    Our result is broadly consistent with the current models of gas enrichment
    and metal loss. However, a more natural explanation for our findings is a
    local relation between the gas-phase metallicity and escape velocity.

    \end{abstract}

  \begin{keywords}

    galaxies: abundances --- galaxies: evolution --- galaxies: fundamental parameters

    \end{keywords}


\section{Introduction}\label{intro}

The abundances of elements heavier than helium in stars and in the interstellar
medium (metallicity) evolve over cosmic time alongside the host galaxies. The
link between chemical evolution and galaxy formation is engraved in the
empirical correlations between metallicity and other physical properties of
galaxies, such as total mass \citep{lequeux+1979, garnett+shields1987,
vila-costas+edmunds1992}.

The advent of large, single-fibre spectroscopic surveys allowed the study
of the systematic variation of metallicity for statistically significant samples
\citetext{Sloan Digital Sky Survey (SDSS), \citealp{york+2000}; 6dF Galaxy Survey,
\citealp{jones+2004}}.
For star-forming galaxies, the gas-phase metallicity averaged over a fibre
aperture was found to correlate tightly with stellar mass 
\citetext{\citealp{tremonti+2004}; hereafter: \citetalias{tremonti+2004}}.
These authors determined the gas-phase metallicity using strong emission lines
in a sample of galaxies drawn from the SDSS Data Release 2 \citep{abazajian+2004}.
For passive galaxies, it is typically easier to measure the metallicity of the
stars, usually by comparing a set of absorption line indices \citep[Lick index
system;][]{worthey+1994} to a grid of models \citep[e.g.][]{worthey1994,
thomas+2010}. The stellar metallicity measured within the SDSS fibres correlates
with both the stellar mass of the galaxy and the root mean
square velocity of the line-of-sight velocity distribution \citep[$\sigma$;][]{thomas+2010}.

Alongside these aperture-averaged relations, metallicity is known to vary systematically
within each galaxy: the existence of radial metallicity gradients has been known
for more than two decades, for both the gas-phase metallicity
\citep[e.g.][]{zaritsky+1994} and the stellar metallicity
\citep[e.g.][]{davies+1993}.
More recently, we have been able to study the resolved metallicity relations with
unprecedented detail and with much larger samples of galaxies, thanks to new large
integral-field spectroscopy surveys \citetext{IFS; SAURON, \citealp{dezeeuw+2002},
ATLAS$^\mathrm{3D}$, \citealp{cappellari+2011}, CALIFA, \citealp{sanchez+2012},
SAMI, \citealp{croom+2012} and MaNGA, \citealp{bundy+2015}}.
For the gas-phase metallicity, the aperture-averaged mass-metallicity relation is
mirrored by two local relations: the existence of a universal radial gradient
\citep{sanchez+2014} and the relation between local metallicity and local
stellar-mass surface density \citep{rosales-ortega+2012, moran+2012, sanchez+2013,
barrera-ballesteros+2016}. The relation with stellar-mass surface density appears
tighter than the physically-driven relations with gas fraction or escape velocity
\citetext{\vesc; \citet{barrera-ballesteros+2018}; hereafter:
\citetalias{barrera-ballesteros+2018}}.

For stellar metallicity, the best aperture-averaged predictor is related to the
gravitational potential, rather than either stellar mass or surface mass
density, as determined using a comparative analysis
\citetext{\citealp{barone+2018}; hereafter: \citetalias{barone+2018}}.
\citet{li+2018} also find that stellar metallicity is approximately constant
along lines of constant $\sigma$, in agreement
with the quantitative analysis of \citetalias{barone+2018}. Although
there is no comparable study of the resolved relations for stellar metallicity,
the \Mgb absorption
index follows a tight relation with \vesc \citep{emsellem+1996}.
This relation is both global and local, i.e. the trend between the local \Mgb
and \vesc is the same across all galaxies, as well as for the values of \Mgb and
\vesc measured inside one effective radius \citep[\re;][]{scott+2009}.\\

Determining whether stellar and gas-phase metallicity follow  different physical
properties of galaxies (and if so, why) is crucial to our understanding of how
galaxies form and evolve. The fact that we cannot state \textit{a priori} what
is the best predictor of gas-phase metallicity demonstrates that the physical
origin of the metallicity relations is still not fully understood. Even less
clear is why stellar metallicity correlates with \vesc whereas previous studies
of the metallicity of star-forming gas propose primary correlations with stellar
mass (or surface mass density).

In order to address this problem, we conduct the first comparative analysis of
the relation between gas-phase metallicity and three structural properties of
galaxies: mass (\M), gravitational potential (\Pot) and surface mass density
(\Sig). We use a volume-limited sample of single-fibre spectroscopy data drawn
from SDSS; in order to
overcome aperture bias inherent to single-fibre data, we introduce the concept of
aperture-matched subsampling: considering subsets of galaxies with a given physical
size (in units of \re) equal to the aperture of the SDSS fibres. Using this method, we find that the
best predictor of gas-phase metallicity is \Pot, in agreement with the results
for stellar metallicity (which were derived using IFS).\\

This work is organised as follows: after introducing the data and the sample
(\refsec{ds}), we present the results of our analysis (\refsec{res}) and a
discussion of the implications (\refsec{disc}); we conclude with a concise
summary of our findings (\refsec{conc}).
Throughout our work, we adopted a $\Lambda\mathrm{CDM}$ cosmology with
$H_0 = 70 \, \mathrm{km\,s^{-1}\,Mpc^{-1}}$, $\Omega_m = 0.3$ and
$\Omega_\Lambda = 0.7$.


\section{Data and Sample}\label{ds}

  We use publicly available data and measurements based on the SDSS Data Release
  7 \citep[SDSS DR7;][]{abazajian+2009}. Throughout this work, unless otherwise
  specified, we use spectroscopic redshifts ($z$), stellar masses (\Mstar),
  star-formation rates (\SFR), specific SFR (\sSFR) and emission line fluxes
  obtained from the publicly available SDSS DR7 MPA/JHU catalogue
  \citep{kauffmann+2003a, brinchmann+2004, salim+2007}.
  The uncertainties on \Mstar and \SFR were calculated as the semi-difference
  between the 84\textsuperscript{th} and 16\textsuperscript{th} percentiles in
  the posterior distribution of each parameter. In addition, we use half-light
  radii (\re) from the NYU Value Added Catalogue \citep[NYUVAC;][]{blanton+2005a}.

  We assume uniform uncertainties on \logre equal
  to $0.05 \; \mathrm{dex}$, determined by comparing \logre to the corresponding
  values measured using the Multi Gaussian Expansion fitting technique
  \citep[MGE, ][]{emsellem+1994}, as implemented by \citet{cappellari2002}. $0.05
  \; \mathrm{dex}$ is obtained from $\mathrm{RMS}/\sqrt{2}$, where RMS is the observed
  root mean square about the best fit linear relation between the NYUVAC and MGE
  values of \logre.

  The conversion between apparent and physical
  size is performed using the redshift-determined angular diameter distance in
  the adopted cosmology (\refsec{intro}). We assume that peculiar velocities are
  negligible, because our sample consists primarily of field and group galaxies:
  at the lowest redshift ($z=0.01$), the error in distance due to a peculiar
  velocity of $300 \, \mathrm{km \, s^{-1}}$ is $\approx 10\%$ or $0.04 \;
  \mathrm{dex}$, less than the measurement uncertainty on \re.

  For each galaxy, we use \Mstar as a proxy for the total mass \M; we further define
  a proxy for the gravitational potential at one effective radius: $\Phi \equiv
  \mathrm{M_*/R_e}$ and a proxy for the average surface mass density within one
  effective radius: $\Sigma \equiv
  \mathrm{M_*/R_e^2}$. These definitions are only first-order approximations,
  that ignore variations both in the intrinsic shape and in
  the mass-to-light ratio of galaxies \citep[e.g.][]{cappellari+2013a}.
  Nevertheless, it has been shown that the photometric proxies adopted here are
  precise enough for a comparative analysis of the stellar population properties,
  with results in excellent qualitative agreement with the equivalent spectroscopic
  proxies \citepalias[i.e. $\mathrm{M_{spec} \equiv \sigma^2 \, R_e}$,
  $\mathrm{\Phi_{spec} \equiv M_{spec} / R_e}$ and
  $\mathrm{\Sigma_{spec} \equiv M_{spec} / R_e^2}$;][]{barone+2018}.

  \subsection{Sample selection}\label{ds.samplesel}

  \begin{figure*}
    \centering
    \includegraphics[type=pdf,ext=.pdf,read=.pdf,width=0.95\textwidth]{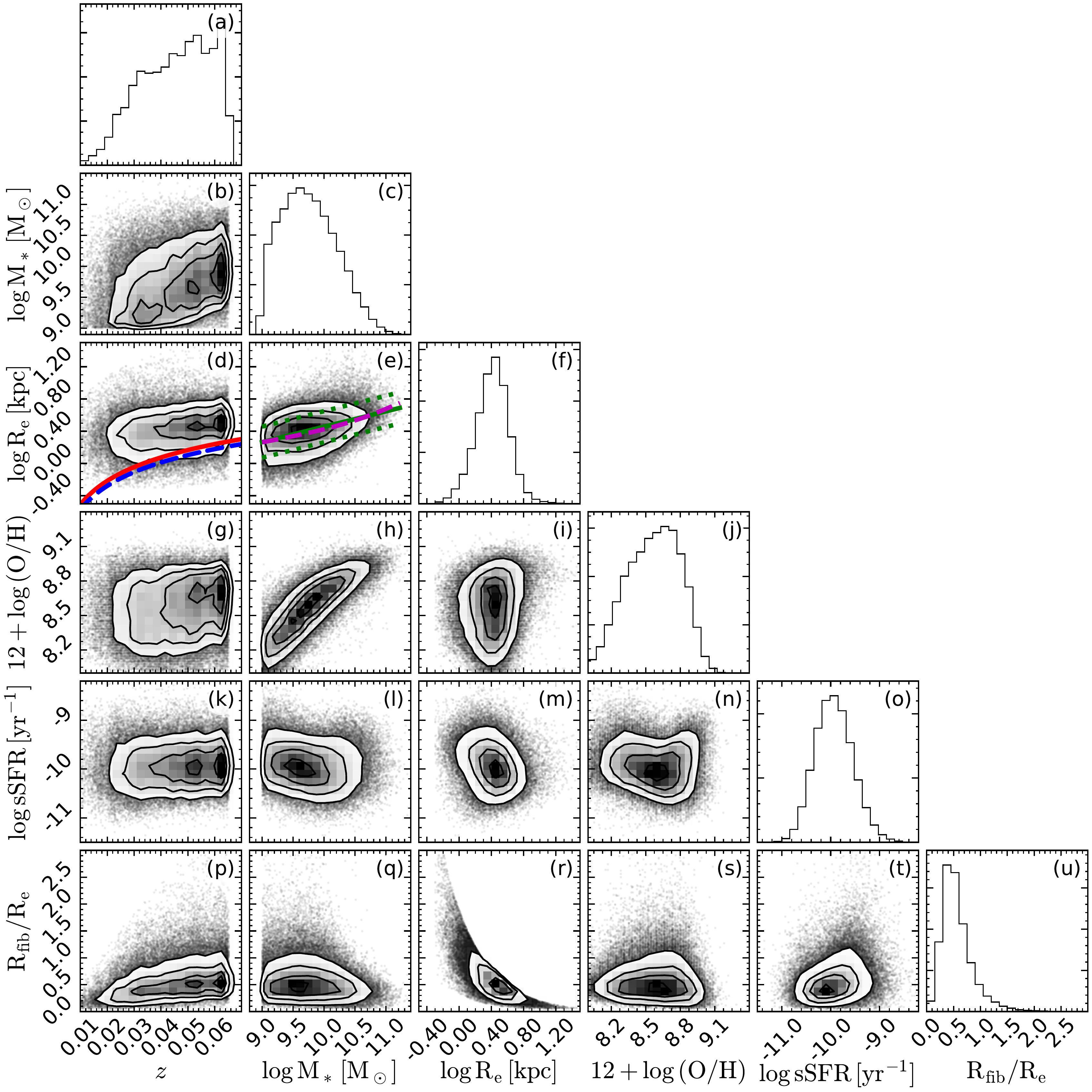}
    {
     \phantomsubcaption\label{f.corn.a}
     \phantomsubcaption\label{f.corn.b}
     \phantomsubcaption\label{f.corn.c}
     \phantomsubcaption\label{f.corn.d}
     \phantomsubcaption\label{f.corn.e}
     \phantomsubcaption\label{f.corn.f}
     \phantomsubcaption\label{f.corn.g}
     \phantomsubcaption\label{f.corn.h}
     \phantomsubcaption\label{f.corn.i}
     \phantomsubcaption\label{f.corn.j}
     \phantomsubcaption\label{f.corn.k}
     \phantomsubcaption\label{f.corn.l}
     \phantomsubcaption\label{f.corn.m}
     \phantomsubcaption\label{f.corn.n}
     \phantomsubcaption\label{f.corn.o}
     \phantomsubcaption\label{f.corn.p}
     \phantomsubcaption\label{f.corn.q}
     \phantomsubcaption\label{f.corn.r}
     \phantomsubcaption\label{f.corn.s}
     \phantomsubcaption\label{f.corn.t}
     \phantomsubcaption\label{f.corn.u}}
    \caption{Summary of the physical properties of our sample. The solid black contours
    enclose the 11\textsuperscript{th}, 39\textsuperscript{th}, 68\textsuperscript{th} and 86\textsuperscript{th} percentiles.
    The solid red line
    in panel~\subref{f.corn.d} shows the radius of the SDSS fibre as a function
    of redshift, while the dashed blue line is the median full width half maximum
    of the $r'$-band point spread function.
    The mass-size relation is highlighted by the solid green line
    (panel~\subref{f.corn.e}; the dotted green lines are offset by the root mean
    square). The dashed magenta line is the mass-size relation from \citep{shen+2003}.
    There is a sharp contrast between
    the tight correlation observed in the mass-metallicity plane
    (panel~\subref{f.corn.h}) and the lack of correlation between size and
    metallicity (panel~\subref{f.corn.i}).
    Our sample has a median fibre coverage
    $\mathrm{R_{fib}/R_e} = 0.5$ (panel~\subref{f.corn.u}).}\label{f.corn}
  \end{figure*}

  In order to guarantee that our results are representative, our main selection
  criteria are designed to build a volume-limited sample, with redshift $0.01
  \leq z \leq 0.065$ and stellar masses $M_* > 10^9 \, \mathrm{M_\odot}$.
  Compared to the current literature, our selection encompasses a smaller and
  closer volume: other studies have used $0.005 < z < 0.25$ \citepalias{tremonti+2004}
  or $0.07 < z < 0.3$ \citep{telford+2016}. Given that our study focuses on the
  importance of fibre coverage, the redshift range is the most important selection
  criterion. However, our results are unchanged if we drop the volume-limited
  condition to adopt either of the two samples cited (\refapp{app.a}).
  The precise limits of the volume-limited sample do not affect our results: we
  repeated our test with $z < 0.08$ and $M_* > 10^{9.5} \, \mathrm{M_\odot}$, and
  $z < 0.11$ and $M_* > 10^{10} \, \mathrm{M_\odot}$, and found the results to be
  qualitatively unchanged.

  Our quality cuts follow the criteria of \citet{mannucci+2010} and
  \citet{telford+2016}: we require a signal-to-noise ratio (SNR) of at least 25
  in the \Ha line, of 5 in the \Hb line and of 3 in both the \SII{6717} and
  \SII{6731} lines (emission-line wavelengths are in \si{\angstrom} unless
  otherwise specified). These criteria ensure that our sample is not biased in
  metallicity \citep[see ][their fig.~1]{telford+2016}. We reject galaxies
  contaminated by active galactic nuclei emission (AGN) using the classification
  of \citet{kauffmann+2003c}, which relies on the BPT diagram
  \citep{baldwin+1981}. The precise threshold adopted to reject AGNs does not
  affect our results, as expected from other studies \citep{kewley+ellison2008}.
  We further impose a lower limit on the equivalent width of the \Ha emission
  line, $\mathrm{EW(H\alpha)}$, in order to remove galaxies where the emission
  lines arise from gas ionised by low-mass, evolved stars rather than OB stars
  associated with star formation \citep{stasinska+2008}. The specific value of the
  minimum $\mathrm{EW(H\alpha)}$ does not affect our analysis: even adopting the most
  conservative selection found in the literature, our results are qualitatively
  unchanged \citep[$\mathrm{EW(H\alpha)} > \SI{14}{\angstrom}$]{lacerda+2018}. However,
  because this strict selection removes approximately 25\% of our sample, we adopt
  the less conservative estimate of \citet{stasinska+2008}, $\mathrm{EW(H\alpha)}
  > \SI{3}{\angstrom}$, as also suggested by other studies
  \citetext{\citealp{davies+2014}, \citealp{cidfernandes+2011}; 17 galaxies do
  not meet this threshold}.
  Our final sample consists of 68959 unique galaxies. This is what we call the
  ``parent sample''.

  \subsection{Metallicity measurements}\label{ds.metal}

  The emission-line fluxes are dust-extinction corrected using Balmer decrement
  $\mathrm{H\alpha/H\beta}$. We assume the \citet{cardelli+1989} dust extinction law
  and case B recombination \citep[$\mathrm{H\alpha/H\beta}=2.86$ for
  $T_e=10^4$ K][]{osterbrock+ferland2006}. We measured the metallicities using the
  new metallicity calibrations proposed by
  \citetext{\citet{dopita+2016}; hereafter: \citetalias{dopita+2016}}. This
  calibration is based on a new grid of photoionization models from the
  MAPPINGS V code (Sutherland et al. in prep.). The combination
  of {\nii/\siii} and {\nii/\ha} provides a metallicity diagnostic that is robust
  against dust reddening and ionisation effects.
  Throughout the paper, the gas-phase metallicity refers always to the
  values calibrated following \citetalias{dopita+2016}, unless otherwise specified.

  In order to explore the effect of different metallicity calibrations, we repeated
  our analysis with three different calibrations: a) the metallicity measurements
  from \citetalias{tremonti+2004}, based on the R23 line ratios ($\mathrm{R23 \equiv
  ([OII]\lambda3727 + [OIII]\lambda\lambda4959,5007)/H\beta}$) and publicly available
  through the MPA/JHU catalogue (\OHT04); b) the metallicity
  calibrated using the O3N2 ratio \citep[\OHO3N2;][]{marino+2013}
  and c) the metallicity calibrated using the N2 ratio \citep[\OHN2;][]{marino+2013}.
  The different calibrations yield the same qualitative results, as reported in
  \refsec{res.ohcalib}.

  \subsection{Sample properties}\label{ds.sprop}

\begin{figure}
    \centering
    \includegraphics[type=pdf,ext=.pdf,read=.pdf,width=1.0\columnwidth]{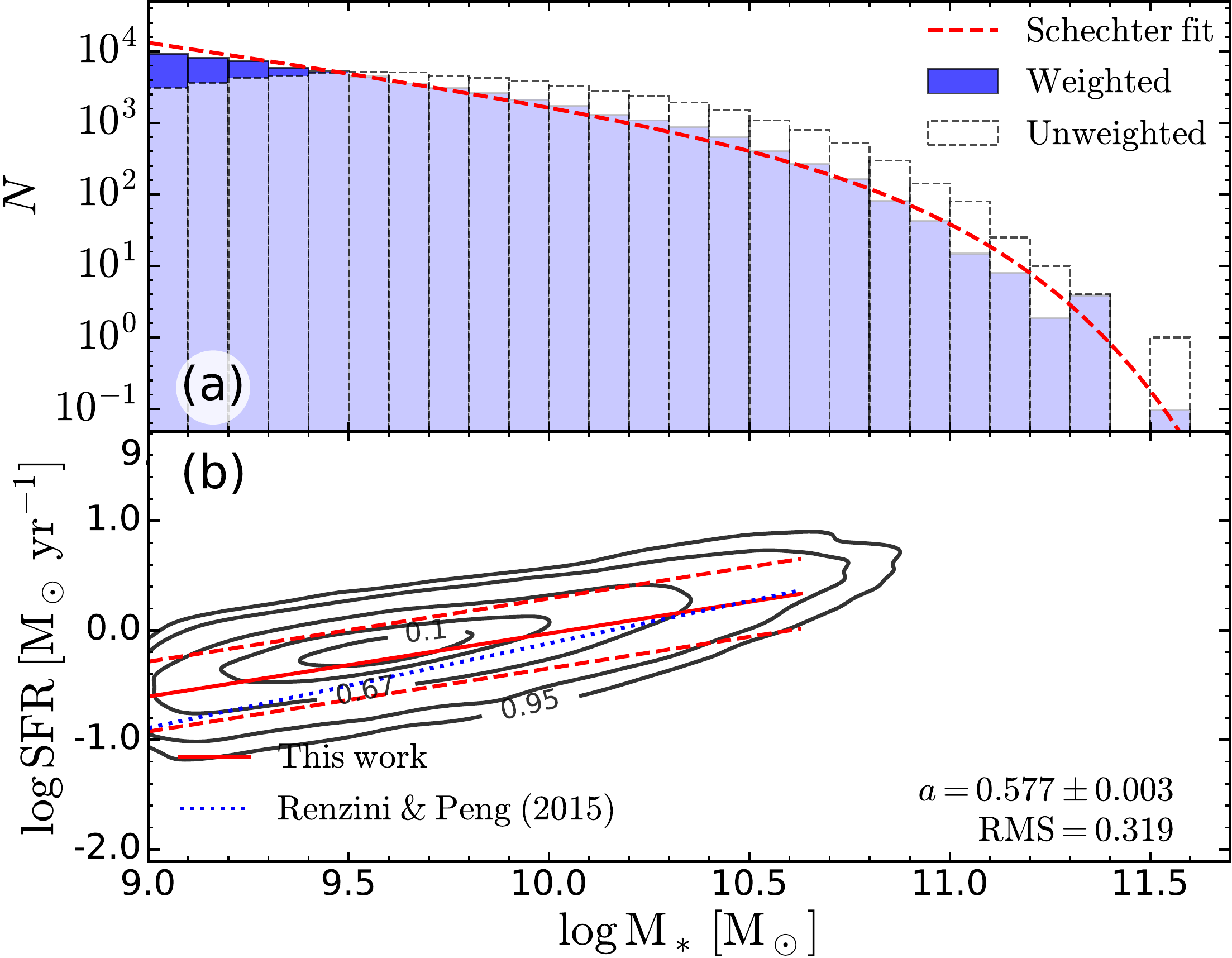}
    {
     \phantomsubcaption\label{f.sample.mass.a}
     \phantomsubcaption\label{f.sample.mass.b}}
    \caption{The mass function of the parent sample is compared to the mass function
    measured in the local Universe \citep[$z < 0.065$;][dashed red line;
    panel~\subref{f.sample.mass.a}]{kelvin+2014}.
    The star-forming main sequence (SFMS) of the parent sample is shown in
    panel~\subref{f.sample.mass.b}; the solid black contours encompass the
    95\textsuperscript{th}, 90\textsuperscript{th}, 67\textsuperscript{th}, 30\textsuperscript{th} and 10\textsuperscript{th} percentiles of the data distribution. The
    best-fit line has slope $a=0.581 \pm 0.003$ (solid red line) and
    $\mathrm{RMS}=0.322 \, \mathrm{dex}$ (dashed red lines). The approximate
    location of the ridge-line of the volume-weighted SFMS measured by
    \citet{renzini+peng2015} is depicted by the dotted blue line.}\label{f.sample.mass}
    \end{figure}

  The properties of the sample are summarised in \reffig{f.corn}: the
  redshift-mass plane and the mass distribution (panels~\subref{f.corn.b}
  and \subref{f.corn.c}) show that there is some incompleteness below
  $\mathrm{M_*} = 10^{9.5} \, \mathrm{M_\odot}$, due to the SNR requirements.
  To address
  this issue we use $V_\mathrm{max}$ weighting: the mass function is shown again
  in \reffig{f.sample.mass.a}, where the dashed histogram is the mass function of
  our sample, the filled blue histogram is the $V_\mathrm{max}$-weighted mass
  function, and the dashed red line is the best-fit Schechter function. We find
  $\log \, \mathcal{M^*} [\mathrm{M_\odot}] = 10.61$ and $\alpha = -1.58$
  \citep[where $\mathcal{M^*}$ and $\alpha$ are defined following ][]{kelvin+2014}.
  Within the uncertainties, these values are consistent with the parameters of
  the Schechter function of disk-dominated galaxies \citep[][their Table~3]{kelvin+2014}.
  The difference between the empty dashed and blue filled histograms highlights
  that the SNR requirements on the emission lines affect our sample selection,
  but thanks to our selection criteria, the correction to the mass function
  is small: repeating our analysis with uniform weighting leaves our results
  qualitatively unchanged.

  Another possible bias that SNR cuts might introduce is against low \sSFR
  galaxies: at fixed mass, galaxies with high \sSFR are more likely to meet the
  quality cut than low \sSFR galaxies. In
  order to assess the importance of this bias, we study the star-forming
  main sequence \citep[SFMS;][]{noeske+2007, renzini+peng2015}.
  We fit a straight line to the
  SFMS, using the least trimmed squares algorithm \citep{rousseeuw+leroy1987,
  rousseeuw+driessen2006} in the free implementation \texttt{lts\_linefit}
  \citep{cappellari+2013a}. We find a slope $a = 0.581 \pm 0.003$, with
  $\mathrm{RMS} = 0.322 \, \mathrm{dex}$ (Fig.~\ref{f.sample.mass.b}, solid red
  line). The general shape of the relation is correctly reproduced
  (\reffig{f.sample.mass.b}), however we notice that our best-fit slope is
  flatter than the ridge line shown in \citet[][
  dotted blue line in \reffig{f.sample.mass.b}]{renzini+peng2015}. This
  discrepancy is partly due to aperture bias at the high-mass end. Because of
  our redshift range, we probe only the innermost region of the highest mass
  galaxies (on average, the largest). For these galaxies, we may
  measure a lower-than-average \SFR. However this bias does not affect our
  results, because we concentrate here on aperture-matched sampling (see
  \refsec{ds.a-mss}). For instance, by repeating the fit with the
  aperture-matched subsample with $\mathrm{R_{fib}/R_e} = 1.5$ we find a
  steeper best-fit slope of the SFMS $a = 0.62 \pm 0.01$.

  We are mainly interested in the relation between three galaxy observables:
  mass, size and metallicity; \reffig{f.corn} provides a summary of the
  distribution of these quantities for our sample. Notice the correlation between
  size and mass (panel~\subref{f.corn.e}), the mass-metallicity relation
  (panel~\subref{f.corn.h}) and the lack of correlation between metallicity and size
  (panel~\subref{f.corn.i}). We use \texttt{lts\_linefit} to fit a straight line
  to the mass-size relation (panel~\subref{f.corn.e}) and find:
  \begin{equation}
      \log \, \mathrm{R_e \; [kpc]} = (0.193\pm0.002) \log \mathrm{M_* \; [M_\odot]} - (1.478 \pm 0.016)
  \end{equation}
  with $\mathrm{RMS} = 0.195$; the best-fit line is depicted as a solid green line
  in \reffig{f.corn.e}, the dashed green lines are offset by $\pm\mathrm{RMS}$.
  Our results are in statistical agreement with the
  literature: \citet{shen+2003} studied the mass-size relation for a
  sample of SDSS late-type galaxies \citep[defined as having
  S\'{e}rsic index $n < 2.5$;][]{sersic1968, shen+2003}. They fit a
  relation of the form:
  \begin{equation}
      \mathrm{R_e \; [kpc]} = \gamma \left(\displaystyle\frac{M}{M_\odot}\right)^\alpha\left(1 + \displaystyle\frac{M}{M_0}\right)^{\beta-\alpha}
  \end{equation}
  We used an uncertainty weighted least-squares algorithm to fit this function
  to our data, and find $\alpha = 0.15 \pm 0.01, \; \beta = 0.57 \pm 0.28, \;
  \gamma = 0.09 \pm 0.01$ and $M_0 = (1.2 \pm 1.1) \times 10^{11} \,
  \mathrm{M_\odot}$, in agreement with the values in the literature
  \citep[$\alpha = 0.14, \; \beta = 0.39, \; \gamma = 0.10$ and $\M_0 = 3.98
  \times 10^{10} \, \mathrm{M_\odot}$; ][their Table~1]{shen+2003}.
  The best-fit relation from \citet{shen+2003} is the dashed green line in
  \reffig{f.corn.e}; our best-fit for the same function overlaps and is not
  depicted for clarity.
  We notice that the parameters $\beta$ and $M_0$ in our fit are unconstrained,
  because of the large uncertainties on the best-fit values: this fact is a
  consequence of the relative lack of massive galaxies in our sample compared to
  \citet{shen+2003}, and emphasizes that a linear fit in log-space is adequate
  to model the mass-size relation for the range of masses and sizes considered in
  this work.

  \subsection{Aperture-matched subsamples}\label{ds.a-mss}

  In order to establish the effect of metallicity gradients on the mass-metallicity
  relation, we divide the volume-limited sample in four subsets, based on the
  ratio between the \textit{apparent} galaxy size and the fibre radius:
  $R_\mathrm{fib}/R_e = 0.5$, $R_\mathrm{fib}/R_e = 1$, $R_\mathrm{fib}/R_e = 1.5$
  and $R_\mathrm{fib}/R_e = 2$. Our results hold for apertures up to
  $R_\mathrm{fib}/R_e \sim 2$, however beyond this limit the
  sample size is too small ($<1000$ galaxies) and our results are not significant.
  The smallest aperture is constrained by the
  requirement that we measure a representative fraction of the galaxy surface
  area. For an exponential light profile, an aperture of radius $0.5 \,
  \mathrm{R_e}$ contains a fraction of the total light equal to $0.21$: below
  this critical fraction the measured gas-phase metallicity does not necessarily
  represent the global metallicity of the galaxy \citep{kewley+2005}. 
  In addition, for apertures smaller than $0.5 \, \mathrm{R_e}$ our proxies for
  the potential
  \Pot and surface mass density \Sig become inadequate: for example the relevant
  potential \Pot cannot be estimated as \MstarRe, but we have to use the ratio
  between the stellar mass enclosed in the fibre ($\mathrm{M_{*,fib}}$) and the
  fibre radius ($\mathrm{R_{fib}}$, expressed in physical units). For these
  reasons, we do not consider apertures smaller than $0.5 \, \mathrm{R_e}$.

  The median fibre coverage in our sample is $\mathrm{R_{fib}/R_e} = 0.5$
  (\reffig{f.corn.u}), with an extended tail to $\mathrm{R_{fib}/R_e} = 2$.
  In order to increase
  the size of each subsample, we introduce a tolerance factor $t=0.13$: for a
  given ratio $\mathrm{R_{fib}/R_e} = r$, we select the corresponding subsample
  using the inequalities $(1 - t) \, r < \mathrm{R_{fib}/R_e} < (1 + t) \, r$. The
  value $t=0.13$ guarantees that each of the aperture-matched samples contains the
  maximum number of galaxies, without overlapping. We repeated the analysis with
  $t=0.10$ and $t=0.05$ and find that the results do not depend on the choice of
  $t$. For reference, given the adopted uncertainty on \logre, the relative
  uncertainty on \re is $\approx 0.12$.


\section{Results}\label{res}

Before presenting our main results, we quantify the effect of aperture bias
(\refsec{res.ab}). We then proceed to study the
relation between stellar mass (\Mstar), size (\re) and gas-phase metallicity (\OH)
by considering subsamples of
fixed mass and size (\refsec{res.fix}). Next we show how \OH
is distributed on the mass-size plane (\refsec{res.msp}). Finally we present
a quantitative study of the potential-metallicity relation (\PZR) and show that
it has less scatter and less residual trends with size than both the mass-metallicity
(\MZR) and surface density-metallicity relations (\SZR; \refsec{res.pzr}).
We conclude by showing that our results do not arise due to the effects of sample or
measurement bias (\refsec{res.bias}) and do not depend on the specific metallicity
calibration adopted (\refsec{res.ohcalib}).

\subsection{Aperture bias}\label{res.ab}

In order to study the correlation between gas-phase metallicity and the physical
size of galaxies at fixed mass, it is important to assess and quantify the bias
due to fixed aperture size.
Past studies have focussed on the redshift-dependent effect of aperture bias: at
fixed physical size a galaxy appears smaller with increasing distance, therefore
the constant aperture of SDSS single-fibre spectroscopy probes increasing
fractional areas with increasing redshift. This effect amounts to a maximum
increase in radial coverage  between $\approx 3$ \citep{telford+2016} and
$\approx 40$ \citepalias{tremonti+2004}, depending on the redshift range selected.
At fixed mass and redshift, the SDSS fibre covers a decreasing fraction of the
galaxy light with increasing galaxy size: for the smallest galaxies the fibre
encompasses all the light, whereas for the largest galaxies the fibre covers
only the innermost, relatively metal rich regions. This observational effect
induces an artificial correlation between size and metallicity at fixed mass and
redshift.

  \begin{figure}
    \centering
    \includegraphics[type=pdf,ext=.pdf,read=.pdf,width=1.0\columnwidth]{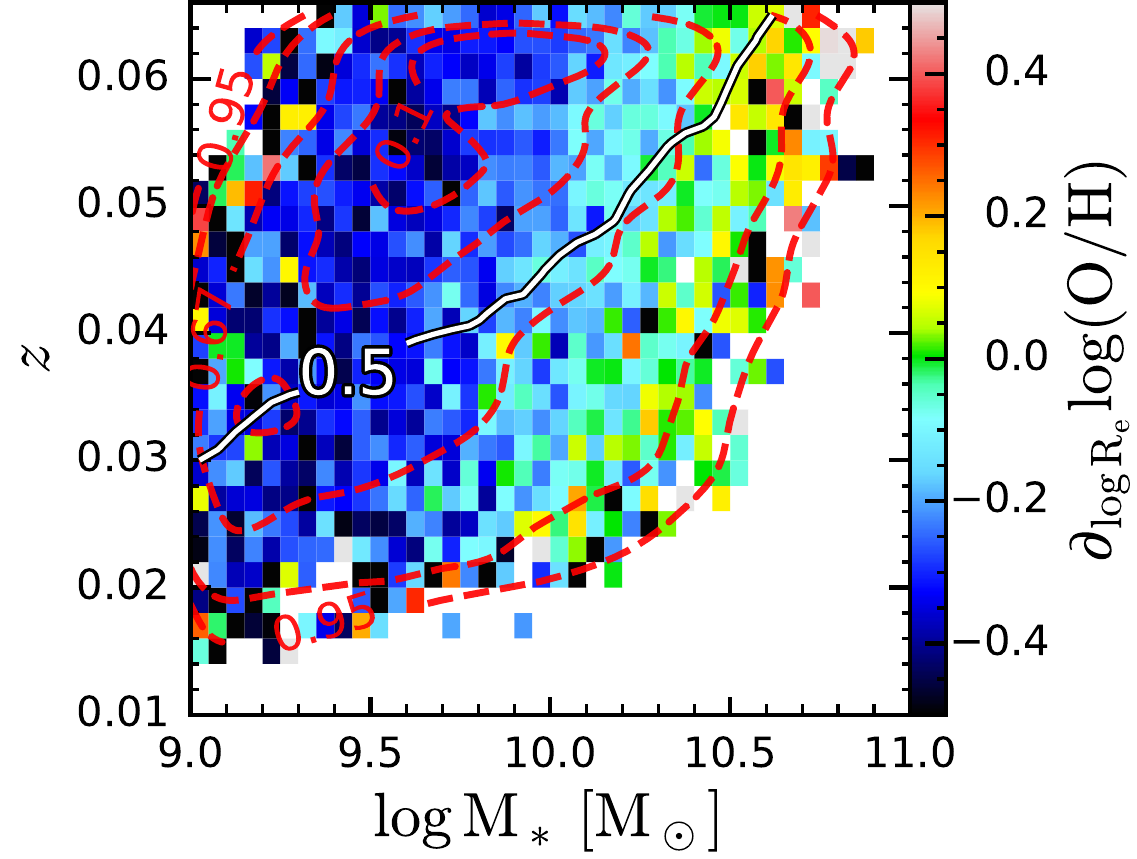}
    \caption{We find a negative metallicity gradient with galaxy size
    ($\partial_{\log \, \mathrm{R_e}} \log \, \mathrm{(O/H)} < 0$); given that
    the gradient due to aperture bias is positive, the negative gradients must
    be physical.
    For each bin at fixed mass and redshift, the colour map indicates the value
    of the slope of the best-fit linear relation between metallicity and size
    ($\partial_{\log \, \mathrm{R_e}} \log \, \mathrm{(O/H)}$). The dashed red
    contour lines enclose the 95\textsuperscript{th}, 90\textsuperscript{th}, 67\textsuperscript{th}, 30\textsuperscript{th} and 10\textsuperscript{th} percentiles
    of the galaxy distribution, while the solid white curve 
    is the locus where the median fibre coverage is $\mathrm{R_{fib}/R_e} = 0.5$:
    below this line the SDSS fibres might not measure accurate global metallicities.
    }\label{f.grads}
  \end{figure}

However for most of our sample we find that size and metallicity are
anticorrelated: in \reffig{f.grads} we show how the size-metallicity correlation
varies on the mass-redshift plane. For each bin in mass and redshift, we use
\texttt{lts\_linefit} to measure the best-fit linear slope between $\log \,
\mathrm{(O/H)}$ and \logre
($\partial_{\log \, \mathrm{R_e}} \log \, \mathrm{(O/H)}$; represented as a
colour map in \reffig{f.grads}; we omit bins with less than 20 galaxies, which
account for a fraction of the sample of less than $0.05$).
The solid white curve is the locus where the median fibre coverage is
0.5: below this curve the SDSS fibre coverage is insufficient to measure a
representative \OH \citep[see \refsec{ds.a-mss} and][]{kewley+2005}. Aperture
bias will tend to produce a positive gradient with galaxy size, because for the
largest galaxies the SDSS fibre probes only the innermost, most metal-rich
region. Therefore the negative gradients observed in \reffig{f.grads} must be
physical (and indeed must be more negative than observed if residual aperture
bias exists). These negative
gradients demonstrate the physical anticorrelation between size and metallicity,
independent of galaxy mass.

\subsection{Gas-phase metallicity at fixed mass and at fixed radius}\label{res.fix}

\begin{figure}
    \centering
    \includegraphics[type=pdf,ext=.pdf,read=.pdf,width=1.0\columnwidth]{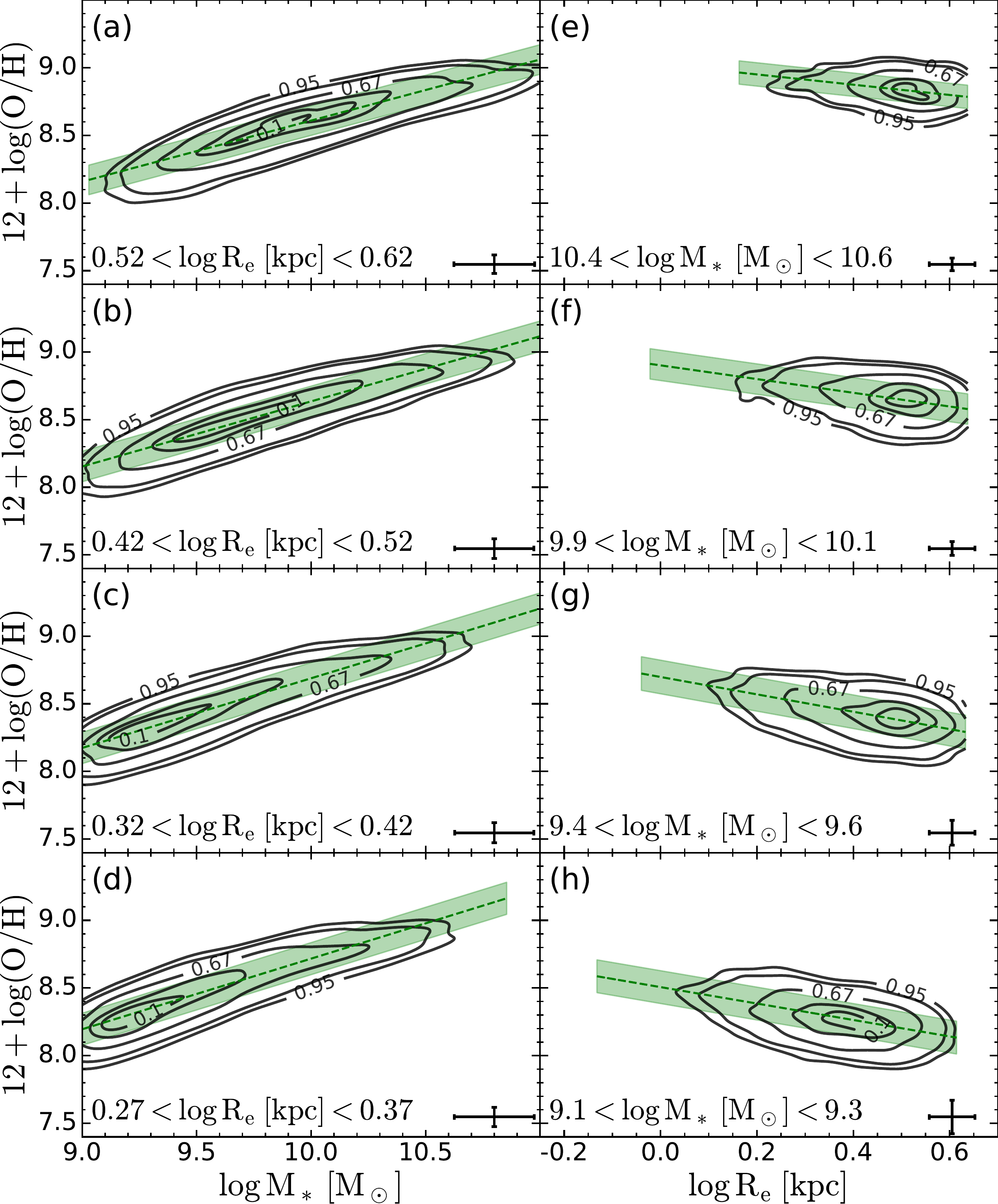}
    {\phantomsubcaption\label{f.fixed.a}
     \phantomsubcaption\label{f.fixed.b}
     \phantomsubcaption\label{f.fixed.c}
     \phantomsubcaption\label{f.fixed.d}
     \phantomsubcaption\label{f.fixed.e}
     \phantomsubcaption\label{f.fixed.f}
     \phantomsubcaption\label{f.fixed.g}
     \phantomsubcaption\label{f.fixed.h}}
    \caption{The dependence of gas-phase metallicity \OH on stellar mass
    \Mstar at fixed size \re (panels~\subref{f.fixed.a}~-~\subref{f.fixed.d})
    and on \re at fixed \Mstar (panels~\subref{f.fixed.e}~-~\subref{f.fixed.h})
    The bins of fixed mass and fixed size are selected along the mass-size
    relation. The contour lines enclose the 95\textsuperscript{th}, 90\textsuperscript{th}, 67\textsuperscript{th}, 30\textsuperscript{th} and 10\textsuperscript{th}
    percentiles of the data distribution, the dashed green lines are the best-fit
    linear relations and the green shaded regions span $\pm \mathrm{RMS}$ about
    the best fit.
    The clear anticorrelation between metallicity and galaxy size at fixed mass
    is physical, because aperture bias would create a positive correlation. The
    best-fit values and RMS for each panel are reported in \reftab{t.fixed},
    where we show that the strength of the mass-metallicity and size-metallicity
    correlations is comparable. The $\mathrm{RMS}$ about the mass-metallicity
    relations is similar to the $\mathrm{RMS}$ about the size-metallicity
    relations, indicating that mass and size are equally good predictors of
    metallicity.}\label{f.fixed}
  \end{figure}

We now study the relation between mass, size and metallicity firstly by
fixing size and letting mass vary, and then by fixing mass and letting size vary. In
order to avoid aperture bias, we consider the aperture-matched subsample with
$\mathrm{R_{fib}/R_e} = 1$, which is equivalent to using IFS data and summing the
spectra over an aperture of radius one \re, an approach widely used in the
literature \citetext{e.g. \citealp{scott+2009}, \citealp{mcdermid+2015},
\citealp{scott+2017}, \citetalias{barone+2018}, \citealp{li+2018}}.
We consider four samples at fixed size
(\subref{f.fixed.a}~-~\subref{f.fixed.d}) and four samples at fixed mass
(\subref{f.fixed.e}~-~\subref{f.fixed.h}); each sample has been centred along the
best-fit mass-size relation (\refsec{ds.sprop}). The width of the mass bins is
$2 \times 0.1 \; \mathrm{dex}$, or twice the measurement uncertainty on $\log \,
\mathrm{M_*}$.
The width of the size bins is $2 \times 0.02 \; \mathrm{dex}$, inferred by
substituting the width of the mass bins in the expression of the best-fit
mass-size relation (this is comparable to the estimated uncertainty on the size
measurements).
The results are shown in \reffig{f.fixed}: in the left column we show the
mass-metallicity relation for four bins at fixed physical size (panels~
\subref{f.fixed.a}~-~\subref{f.fixed.d}); in the right column we show the
size-metallicity relation for four bins at fixed mass (panels~
\subref{f.fixed.e}~-~\subref{f.fixed.h}). The mass or size range of each
subsample is reported in the lower left corner of each panel, as well as in
\reftab{t.fixed} (Column 1).

In each panel, the dashed green line represents the best-fit linear relation and
the green shaded region covers an offset of $\pm \mathrm{RMS}$ in metallicity.
Although a linear relation does not describe accurately the mass-metallicity
relation (the slope flattens at high masses), here our aim is to find the
average gradient, and for this purpose a simple linear fit suffices.
The best-fit slope $a$ and the $\mathrm{RMS}$ about the best-fit line are
reported in \reftab{t.fixed} (Columns 3 and 5). We find everywhere a
statistically significant trend (with significance $a/\sigma_a > 10$;
\reftab{t.fixed} Column 4). The RMS is very similar for the mass-metallicity
and size-metallicity relations - so mass and size are equally good predictors of
metallicity (which, given the scatter about the mass-size relation, would not be
the case if one were a better predictor for metallicity than the other). The
overall change in metallicity $\Delta \log
\mathrm{(O/H)}$ is much smaller for the size-metallicity relation than for the
mass-metallicity relation (Column 6), but this fact does not mean that size
is less important than mass when it comes to metallicity, because 
the dynamic range in size is much smaller than the dynamic range in mass
($\Delta \log \mathrm{R_e} \lesssim 0.6 \; \mathrm{dex}$;
panels~\subref{f.fixed.e}~-~\subref{f.fixed.h} compared to $\Delta \log
\mathrm{M_*} \gtrsim 2 \; \mathrm{dex}$;
panels~\subref{f.fixed.a}~-~\subref{f.fixed.d}). The difference in range between
mass and size arises from three facts: in part, it reflects a real feature of the
galaxies in our sample (Figs~\ref{f.corn.a} and \ref{f.corn.c}: compare the width
of the distributions of mass and size); in part, it results from the constraint
that $\mathrm{R_{fib}/R_e}$ is within a narrow range of unity; and finally, it is
part due to the flat slope of the mass-size relation ($a \approx 0.2$), so that
we need to use a relatively wide range in mass to obtain the full range in size
- which goes against the grain of this test (i.e. letting size vary at fixed
mass).

A \textit{caveat} of this analysis is that even the aperture-averaged
metallicity within one \re~may still suffer from some degree of aperture bias,
because earlier morphological types have more concentrated light profiles
\citep{iglesias-paramo+2016}. In order to measure an unbiased metallicity,
\citet{iglesias-paramo+2016} recommend using an aperture with radius equal to
$2 \; \mathrm{R_e}$. Unfortunately our sample does not possess enough galaxies to repeat this
quantitative analysis with the aperture-matched sample with
$\mathrm{R_{fib}/R_e} = 2$; however we remark that our results are unchanged if
we adopt the aperture-matched sample with $\mathrm{R_{fib}/R_e} = 1.5$. In
addition, if we use only galaxies with identical light profiles (selected by
their measured S{\'e}rsic index), our results are unchanged (see
\refapp{app.b}). We therefore conclude that aperture bias does not play a
determining role in our results.

\begin{table}
  \begin{center}
  \setlength{\tabcolsep}{2pt}
  \caption{$\log \, \mathrm{O/H}$ variation with size at fixed mass and with mass at fixed size.}\label{t.fixed}
  \begin{tabular}{cccccc}
  \hline
  Figure & $\log \mathrm{R_e \; [kpc]}$ & $a_M\pm\sigma_a$ & $a_M/\sigma_a$ & $\mathrm{RMS}$ & $\Delta \log \mathrm{O/H}$ \\
  (1) & (2) & (3) & (4) & (5) & (6) \\
  \hline
 (\ref{f.fixed.a}) & $[0.52; 0.62]$ & $ 0.45\pm0.01$ & 65.52 & 0.12 &  0.89 \\
 (\ref{f.fixed.b}) & $[0.42; 0.52]$ & $ 0.48\pm0.01$ & 82.83 & 0.12 &  1.15 \\
 (\ref{f.fixed.c}) & $[0.32; 0.42]$ & $ 0.52\pm0.01$ & 69.24 & 0.12 &  1.02 \\
 (\ref{f.fixed.d}) & $[0.27; 0.37]$ & $ 0.54\pm0.01$ & 54.32 & 0.13 &  0.99 \\
  \hline
  \\
  \hline
  Figure & $\log \mathrm{M_* \; [M_\odot]}$ & $a_R\pm\sigma_a$ & $a_R/\sigma_a$ & $\mathrm{RMS}$ & $\Delta \log \mathrm{O/H}$ \\
  (1) & (2) & (3) & (4) & (5) & (6) \\
  \hline
 (\ref{f.fixed.e}) & $[10.4; 10.6]$ & $-0.38\pm0.04$ & 10.66 & 0.10 & -0.18 \\
 (\ref{f.fixed.f}) & $[ 9.9; 10.1]$ & $-0.51\pm0.02$ & 21.10 & 0.12 & -0.33 \\
 (\ref{f.fixed.g}) & $[ 9.4;  9.6]$ & $-0.64\pm0.02$ & 31.11 & 0.13 & -0.43 \\
 (\ref{f.fixed.h}) & $[ 9.1;  9.3]$ & $-0.61\pm0.02$ & 27.15 & 0.13 & -0.45 \\
  \hline
  \end{tabular}
  \end{center}

  For each of the panels in \reffig{f.fixed}, we detail: 1) the size
  range (panels~\subref{f.fixed.a}~-\subref{f.fixed.d}) or the mass range
  (panels~\subref{f.fixed.e}~-~\subref{f.fixed.f}); 2) the
  best-fit slope of the relation between $\log \, \mathrm{O/H}$ and
  $\log \, \mathrm{M_*}$
  (panels~\subref{f.fixed.a}~-~\subref{f.fixed.d}) or between
  $\log \, \mathrm{O/H}$ and and \logre
  (panels~\subref{f.fixed.e}~-~\subref{f.fixed.f}); 3) the
  statistical significance of the trend (in units of the standard deviation
  $\sigma_a$); 4) the observed root mean square around the best-fit relation (5)
  and the total variation in $\log \, \mathrm{O/H}$ over the observed range (6).
\end{table}

Despite these limitations, we find sufficient evidence to state that galaxy size
affects the metallicity of the
star-forming gas independent of galaxy mass; the logarithmic slope of the
size-metallicity relation has opposite sign but comparable absolute value to the
logarithmic slope of the mass-metallicity relation ($-1.2 < a_R/a_M < -0.8$;
\reftab{t.fixed}).

\subsection{Gas-phase metallicity on the mass-size plane}\label{res.msp}

\begin{figure}
    \centering
    \includegraphics[type=pdf,ext=.pdf,read=.pdf,width=1.0\columnwidth]{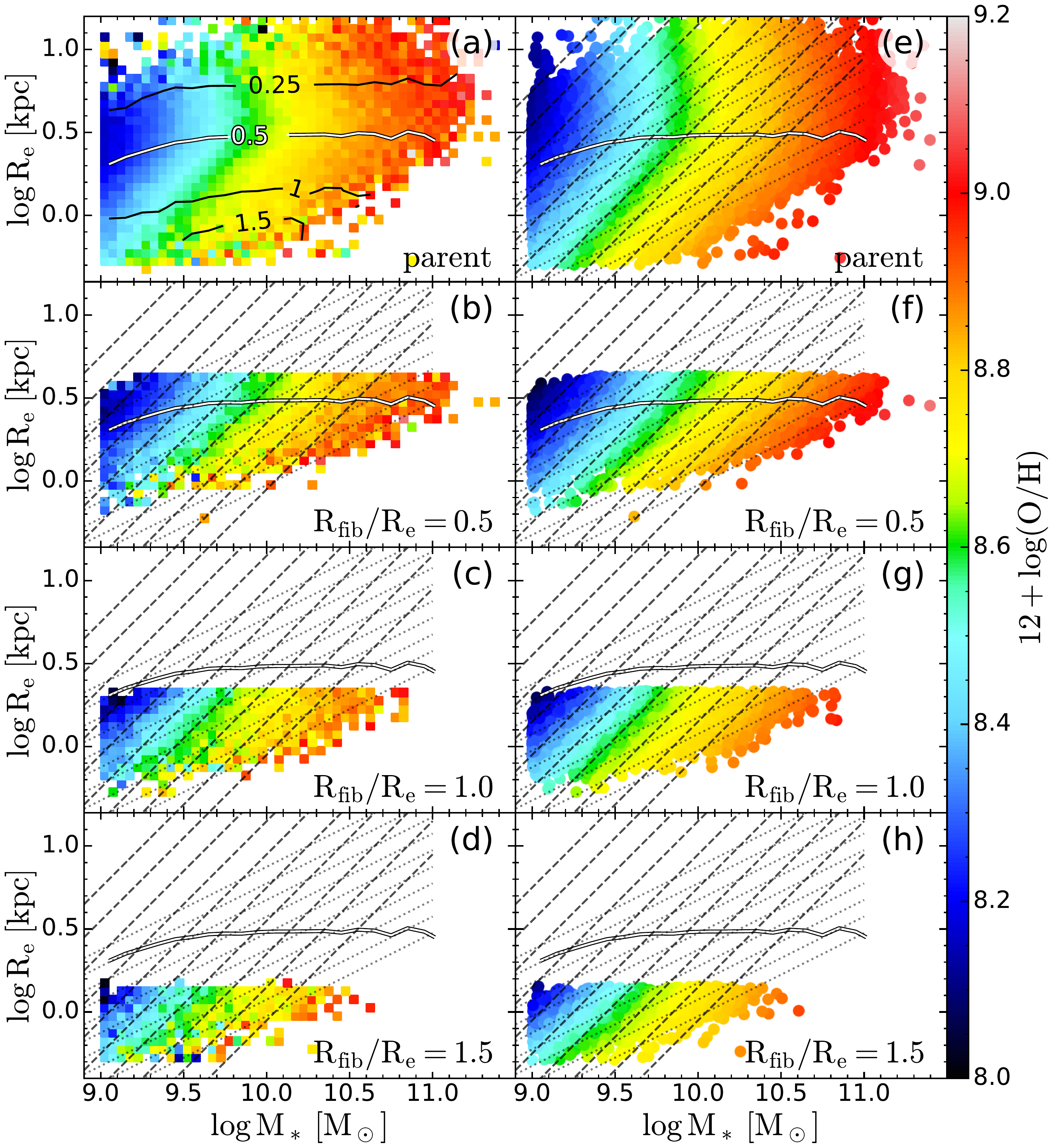}
    {\phantomsubcaption\label{f.masssize.a}
     \phantomsubcaption\label{f.masssize.b}
     \phantomsubcaption\label{f.masssize.c}
     \phantomsubcaption\label{f.masssize.d}
     \phantomsubcaption\label{f.masssize.e}
     \phantomsubcaption\label{f.masssize.f}
     \phantomsubcaption\label{f.masssize.g}
     \phantomsubcaption\label{f.masssize.h}
     \phantomsubcaption\label{f.masssize.i}
     \phantomsubcaption\label{f.masssize.j}}
    \caption{Gas-phase metallicity of the SDSS galaxies on the mass-size plane.
    The first row shows the results for the parent sample, the second, third and
    fourth rows show the aperture-matched subsamples with $\mathrm{R_{fib}/R_e}
    = 0.5$, $\mathrm{R_{fib}/R_e} = 1$ and $\mathrm{R_{fib}/R_e} = 1.5$. The
    panels in the left column show the raw (noisy) data, while the right panels
    shows the underlying distribution, reconstructed using the LOESS algorithm.
    In the top left panel, the solid lines are lines of constant (median) fibre
    coverage: the solid white curve with $\mathrm{R_{fib}/R_e} = 0.5$ (replicated
    in all panels) is the critical line below which the SDSS fibres can
    accurately measure a representative \OH. Above this limit, the lines of constant
    metallicity are approximately
    vertical (panels~\subref{f.masssize.a} and \subref{f.masssize.f}); however
    in this region the galaxies have insufficient fibre coverage. Where the SDSS
    fibre covers at least $0.5 \, \mathrm{R_e}$ (i.e. below the white curve),
    metallicity is constant along lines of constant \Pot (dashed black lines),
    rather than lines of constant mass (vertical) or
    constant \Sig (dotted black lines; see panels~\subref{f.masssize.b}~-~\subref{f.masssize.e}
    and \subref{f.masssize.f}~-~\subref{f.masssize.j})
    }\label{f.masssize}
  \end{figure}

  In \reffig{f.masssize} we show the distribution of \OH on the mass-size plane
  \citep[][]{cappellari+2013b}. Each row displays the results for a
  different sample: the parent sample (top row), and the aperture-matched
  subsamples with $\mathrm{R_{fib}/R_e} = 0.5$ (second row),
  $\mathrm{R_{fib}/R_e} = 1$ (third row), $\mathrm{R_{fib}/R_e} = 1.5$
  (fourth row) and $\mathrm{R_{fib}/R_e} = 2$ (bottom row).
  The left column shows the distribution for the raw data (in
  square bins of $0.05 \, \mathrm{dex}$ side,
  panels~\subref{f.masssize.a}~-~\subref{f.masssize.e}); in the right
  column the data has been smoothed using the two-dimensional locally-weighted
  robust regression technique \citep[LOESS; ][]{cleveland1979,
  cleveland+devlin1988}, as implemented by \citet{cappellari+2013b}; here the
  circles represent individual galaxies, colour-coded by their inferred \OH.
  The solid lines in panel~\subref{f.masssize.a} are lines of constant median fibre
  coverage $\mathrm{R_{fib}/R_e}$. The solid white curve indicates the locus where the
  (median) fibre coverage is $\mathrm{R_{fib}/R_e} = 0.5$; this line is reproduced
  in all the panels.
  The region above this line consists of galaxies where the SDSS coverage is
  insufficient to measure a representative \OH \citep[see \refsec{ds.a-mss} and][]{kewley+2005}.
  In all other panels, the dashed and the dotted black lines are lines of constant \Pot
  and \Sig respectively.

  For galaxies with $\mathrm{R_{fib}/R_e} < 0.5$, we find
  that the lines of constant \OH are approximately vertical, i.e. lines of
  constant mass. However for these galaxies the fractional fibre coverage is
  $\mathrm{R_{fib}/R_e} \approx 0.25$, therefore: (i) the ratio $\mathrm{M_*/R_e}$
  is not a good proxy for \Pot and (ii) the measured metallicity is not
  representative of the galaxy metallicity.
  In fact, the observed trend above the white curve is
  exactly what is expected from aperture bias: at fixed \Mstar, the largest
  galaxies have the lowest fractional fibre coverage, therefore their fibre
  metallicity is higher than the average value for each galaxy. As a result, at fixed \Mstar, the
  largest galaxies are more metal rich than predicted by the local line of
  constant \Pot.

  In contrast, for galaxies with $\mathrm{R_{fib}/R_e} > 0.5$ (i.e. below the
  solid white curve), we find that the lines of constant \OH are approximately
  aligned to lines of constant \Pot.
  This is true for both the full sample (panels~\subref{f.masssize.a} and
  \subref{f.masssize.f}) and for each of the individual aperture-matched samples
  (panels~\subref{f.masssize.b}~-~\subref{f.masssize.e} and
  \subref{f.masssize.g}~-~\subref{f.masssize.j}).
  If we exclude the unreliable region with $\mathrm{R_{fib}/R_e} < 0.5$, we find
  that the lines of constant metallicity: i) have approximately uniform slope
  across the full mass and size range, ii) appear
  to be aligned along lines of constant \Pot and iii) nowhere in the valid
  part of the mass-size plane ($\mathrm{R_{fib}/R_e} > 0.5$) are they aligned
  with lines of constant \M or \Sig.

  We also observe that the zero-point of the metallicity relations within each
  aperture-matched subsample are different: at a given value of \Pot, metallicity
  increases from the sample with $\mathrm{R_{fib}/R_e} = 2$ to the
  sample with $\mathrm{R_{fib}/R_e} = 0.5$. However this change depends on the
  metallicity calibration adopted \citep{kewley+ellison2008}, and may affect the
  relation for the parent sample (see \refsec{res.ohcalib}).

  \subsection{The gas-phase metallicity-potential relation}\label{res.pzr}

  We have seen that the gas-phase metallicity on the mass-size plane is
  approximately constant along
  lines of constant \Pot (\refsec{res.msp}) rather than lines of constant \M or \Sig.
  We now seek to determine which of the three physical proxies \M, \Pot and
  \Sig, is the best predictor of \OH. To this end, we consider the relation between
  \OH and each of \M, \Pot and \Sig, and we look at three metrics: (i) the RMS
  about the running median, (ii) the Spearman rank correlation coefficient
  (\rhosp), and (iii) the presence and strength of trends between the residuals
  with respect to the median and the physical size (quantified using \rhospred,
  the relevant value of the Spearman rank correlation coefficient).

  \begin{figure*}
    \centering
    \includegraphics[type=pdf,ext=.pdf,read=.pdf,width=1.0\textwidth]{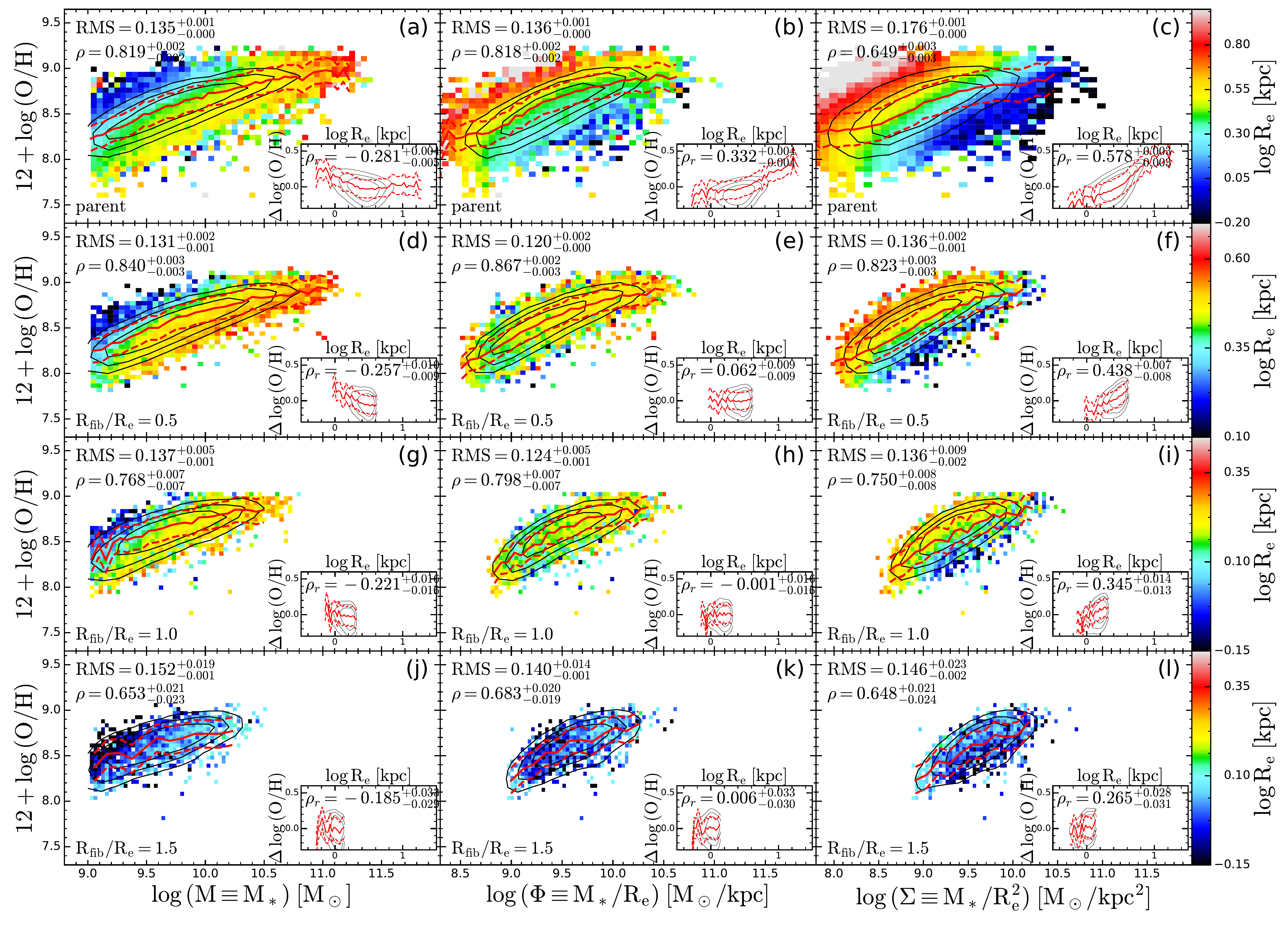}
    {\phantomsubcaption\label{f.oh.a}
     \phantomsubcaption\label{f.oh.b}
     \phantomsubcaption\label{f.oh.c}
     \phantomsubcaption\label{f.oh.d}
     \phantomsubcaption\label{f.oh.e}
     \phantomsubcaption\label{f.oh.f}
     \phantomsubcaption\label{f.oh.g}
     \phantomsubcaption\label{f.oh.h}
     \phantomsubcaption\label{f.oh.i}
     \phantomsubcaption\label{f.oh.j}
     \phantomsubcaption\label{f.oh.k}
     \phantomsubcaption\label{f.oh.l}}
    \caption{Gas-phase metallicity (expressed as $12 + \log \mathrm{O/H}$) versus
    mass (\M; left column), gravitational potential (\Pot; central
    column) and surface mass density (\Sig; right column). The first row
    (panels~\subref{f.oh.a}~-~\subref{f.oh.c}) shows the results for the parent
    sample; the second, third and fourth rows show the results for the
    aperture-matched samples with $\mathrm{R_{fib}/R_e}=0.5$,
    $\mathrm{R_{fib}/R_e}=1$ and $\mathrm{R_{fib}/R_e}=1.5$ respectively.
    The solid black contour lines enclose the 90\textsuperscript{th},
    75\textsuperscript{th}, and 50\textsuperscript{th} percentiles of the galaxy
    distribution. The red solid line is the running volume-weighted median in
    thirty equal-size bins of the physical parameter, the dashed lines are
    the 16\textsuperscript{th} and 84\textsuperscript{th} percentiles. In the
    inset diagram of each panel we show the residuals with respect to the
    running median, as a function of \logre. The galaxies are colour-coded with
    their physical size (\re) to highlight the presence or absence of residual
    trends. For the aperture-matched samples, the \PZR relations have
    less scatter (RMS; top left corner of each panel), higher Spearman rank
    correlation coefficient (\rhosp; bottom left corner) and less residual trends
    with size (lower \rhospred in the inset diagrams) than both the \MZR and \SZR
    relations.
    }\label{f.oh}
  \end{figure*}

  Firstly, we compare the gas-phase metallicity \OH with mass (\Mstar),
  gravitational potential (\Pot) and surface mass density (\Sig) for the full
  sample (Fig.~\ref{f.oh.a}~-~\subref{f.oh.c}). The galaxies are binned in both mass
  and metallicity, with each bin colour-coded by its median effective radius in
  physical units (\re); the solid red line is the volume-weighted median metallicity
  in thirty equal-size bins of the physical parameter, the dashed red lines are the
  16\textsuperscript{th} and 84\textsuperscript{th}
  percentiles. The solid black lines are isodensity contours. In the top left
  corner of each panel we report the RMS, in the bottom left corner we report the
  Spearman rank correlation coefficient (\rhosp). The uncertainties on RMS and
  \rhosp were estimated by bootstrapping 75 percent of the sample 1000 times. 
  All three panels show clear residual trends of \OH with \re, that are visible
  as gradients in the colour hue in the main panels; in addition, these trends
  are quantified in the inset diagrams, where we show the residuals of each \OH
  relation as a function of \logre, and where we report the value of \rhospred,
  the Spearman rank correlation coefficient for
  the distribution in the $\Delta \, \log \, (\mathrm{O/H}) - \log \mathrm{R_e}$
  space.

  For the parent sample (Figs~\ref{f.oh.a} and \ref{f.oh.b}), the \MZR and \PZR
  relations are approximately equivalent, with similar
  values of the RMS ($0.135$ vs. $0.136$) and \rhosp ($0.819$ vs. $0.818$), and
  similar (but opposite) \rhospred for the residuals as a function of \logre
  ($-0.281$ vs. $0.332$). The \SZR relation is clearly worse: it has larger
  RMS ($0.176$), lower \rhosp ($0.649$) and higher \rhospred for the residuals as a
  function of \logre ($0.578$).

  In the third row of \reffig{f.oh} we show the same diagram for the
  aperture-matched subsample with $\mathrm{R_{fib}/R_e}=1$
  (panels~\subref{f.oh.g}-\subref{f.oh.i}).
  The \PZR relation is now better than both the
  \SZR and \MZR correlations: it has higher \rhosp, lower RMS and it has no
  residual trends with galaxy size (\rhospred is $-0.001$ for
  the residuals of the \PZR relation, compared to $-0.221$ and $0.345$ for
  the residuals of the \MZR and \SZR relations).
  Equivalent results are seen also in the aperture-matched subsample with
  $\mathrm{R_{fib}/R_e}=0.5$ (panels~\subref{f.oh.d}~-~\subref{f.oh.f}) and
  $\mathrm{R_{fib}/R_e}=1.5$ (panels~\subref{f.oh.j}~-~\subref{f.oh.l}).
  For apertures larger than
  $\mathrm{R_{fib}/R_e}=1.5$, the relevant aperture-matched samples have
  too few galaxies to discriminate between the three metallicity relations. For
  apertures smaller than $\approx 0.5 \;\mathrm{R_e}$: (i) \MstarRe might
  not be a good proxy for \Pot in such small apertures and (ii) the metallicity
  measured within $\mathrm{R_{fib}} < 0.5 \, \mathrm{R_e}$ is not representative of
  the total galaxy metallicity (see \refsec{res.msp}).

  In conclusion, aperture-matched subsamples: (i) highlight the presence
  of residual size-metallicity trends at fixed mass and surface mass density and
  (ii) demonstrate that \Pot is a slightly better (lower RMS) and less biased
  (less residual with size) predictor of metallicity than either \M or
  \Sig.

\subsection{Measurement bias and correlation between mass and size}\label{res.bias}

One major possible source of bias is the correlation between stellar mass
measurements and fibre size, because \Mstar is also estimated using fibre spectra
and so these too vary with the coverage of the galaxy by the fibre.
For this reason, we repeated our analysis using a purely photometric mass
estimator that is completely independent of $\mathrm{R_{fib}}$
\citep[$\mathrm{M_{*,p}}$;][]{taylor+2011}, we obtain the same results as using
\Mstar: $\mathrm{M_{*,p}/R_e}$ is a better predictor of aperture-averaged metallicity than
either $\mathrm{M_{*,p}}$ alone or $\mathrm{M_{*,p}/R_e^2}$. We can therefore rule out
the possibility that our results are caused by correlated errors between the mass
measurements and the fibre coverage.

We can also exclude the possibility that our results arise from selection bias
or correlated uncertainties. In principle, our SNR constraints (\refsec{ds})
might bias the sample against galaxies with low surface brightness, because a
lower fraction of their light enters the SDSS fibre than for galaxies of equal
total brightness but smaller apparent size.
However, even by selecting our sample subject to $z<0.065$ and $\Sigma_* \geq
10^{8.5} \, \mathrm{M_\odot \, arcsec^{-2}}$ we find results consistent with 
those shown in \refsec{res.pzr}: the \MZR and \SZR relations have clear residual
trends with the physical size of galaxies, both for the full sample and for the
aperture-matched subsamples. Hence we rule out our results being a product of
the selection criteria.

Correlated measurement uncertainties between \Mstar and \re can
artificially reduce the scatter in the \PZR relation compared to the \OH-\M
relation. However the measurements of \Mstar use a different measurement than the
measurements of \re, which rules out correlated random uncertainties. Another
possibility is the presence of
correlated systematic errors due to fitting the galaxy light profiles with a
pre-determined function (in our case, the family of S\'ersic light profiles).
However our results are qualitatively unchanged if we replace the S\'ersic \re
with the MGE \re, which are almost non-parametric \citep{emsellem+1994,
cappellari2002}.

Finally, the physical correlation between \Mstar and \re (\refsec{ds}) could
artificially reduce the scatter in the \PZR relation compared to the \MZR
relation. If we were using the wrong model to fit the empirical relations, we
could be imposing a coherent structure in the residuals, which then combines
with the correlation between \Mstar and \re to induce (or remove) correlations
between the residuals and galaxy size. 
However this possibility is excluded because we use a non-parametric running
median, which does not impose any pre-determined functional form on the
metallicity relations and on the residuals.

It is worth remarking that - by construction - the observational uncertainty
on \MstarRe is larger than the observational uncertainty on \Mstar alone. For
this reason we conclude that whenever the observed RMS about the \PZR
relation is equal to or smaller than the RMS about the \MZR relation, then
the intrinsic scatter in the \PZR relation is necessarily smaller than that
in the \MZR relation (we have ruled out the presence of correlated measurement
uncertainties). The same argument holds for the \SZR relations compared to
both the \MZR and \PZR relations: the fact that the \SZR relation has
larger RMS than both the \MZR and \PZR relations does not imply by itself
that \M or \Sig are better predictors of \OH than \Sig. However the residuals of
the \SZR relation show a clear trend with galaxy size, whereas no such trend
is observed for the residuals of the \PZR relation. This fact suggests that
at least part of the increase in RMS from the \PZR to the \SZR relation
is because \Sig does not correctly take into account the metallicity information
contained in the size of galaxies. We conclude therefore that, of the three
structural parameters $\mathrm{M} \; (\equiv \mathrm{M_*})$, $\Phi \; (\equiv
\mathrm{M_*/R_e})$ and $\Sigma \; (\equiv \mathrm{M_*/R_e^2})$, the best predictor
of \OH is \Pot.

\subsection{Alternative metallicity measurements}\label{res.ohcalib}

The results shown so far use the metallicity calibration of
\citetalias{dopita+2016} (\refsec{ds.metal}). However, adopting a different
metallicity calibration affects the aperture-averaged metallicity measurements inside
a constant aperture \citetext{\citealp{mannucci+2010};~\citetalias{sanchez+2017}}.
For this reason it is important to test whether our results are altered when a
different metallicity calibration is adopted. We repeated our full analysis using three
alternative calibrations: \OHT04, \OHN2 and \OHO3N2 (see again \refsec{ds.metal}).
In general we find very good qualitative agreement between the different
calibrations: the \PZR relation has less observed scatter, higher \rhosp and
weaker residual trends with galaxy size than both the \MZR and \SZR relations,
regardless of the specific metallicity calibration adopted.

  \begin{figure}
    \centering
    \includegraphics[type=pdf,ext=.pdf,read=.pdf,width=1.0\columnwidth]{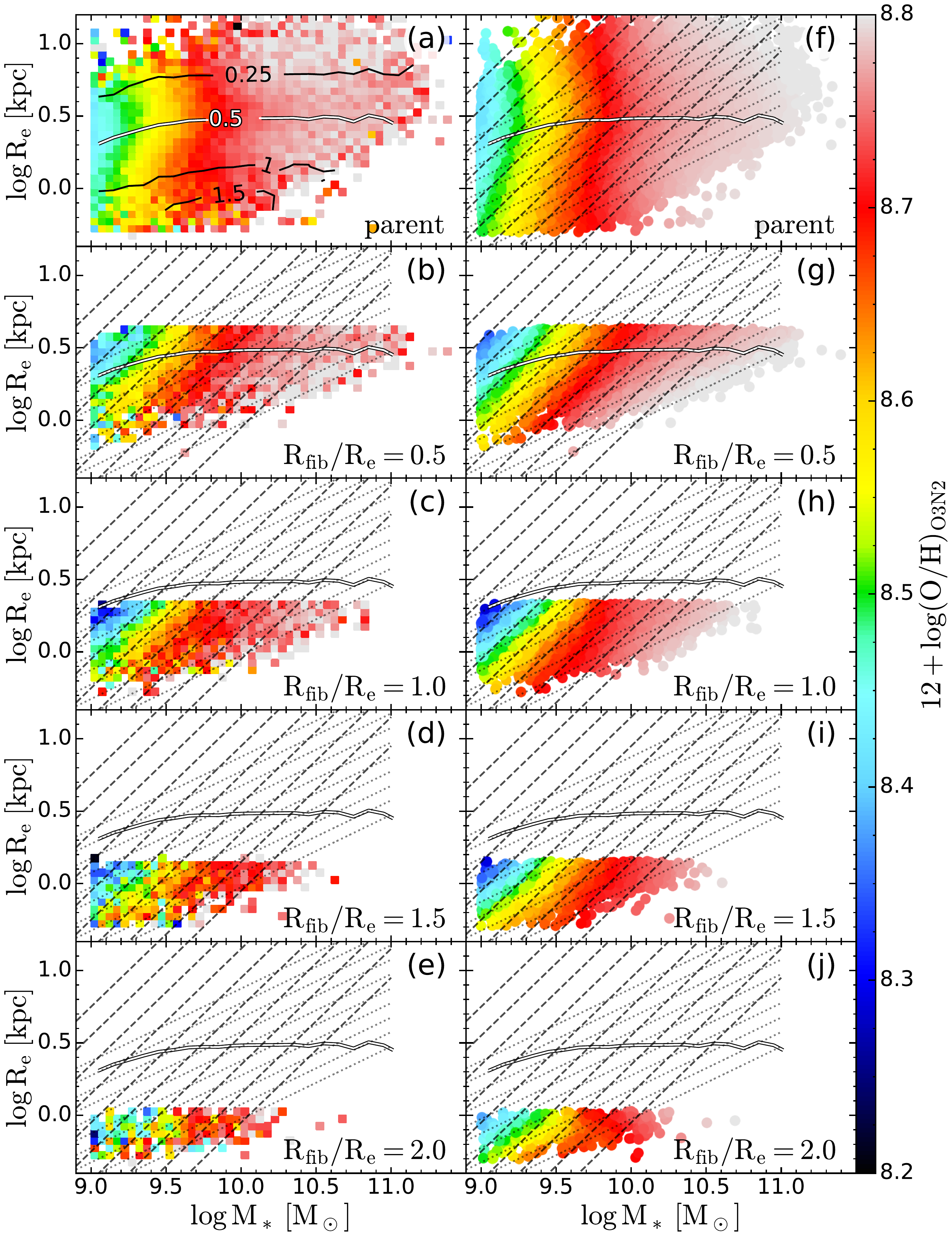}
    {\phantomsubcaption\label{f.masssize.o3n2.a}
     \phantomsubcaption\label{f.masssize.o3n2.b}
     \phantomsubcaption\label{f.masssize.o3n2.c}
     \phantomsubcaption\label{f.masssize.o3n2.d}
     \phantomsubcaption\label{f.masssize.o3n2.e}
     \phantomsubcaption\label{f.masssize.o3n2.f}
     \phantomsubcaption\label{f.masssize.o3n2.g}
     \phantomsubcaption\label{f.masssize.o3n2.h}
     \phantomsubcaption\label{f.masssize.o3n2.i}
     \phantomsubcaption\label{f.masssize.o3n2.j}}
    \caption{Gas-phase metallicity of the SDSS galaxies on the mass-size plane,
    using \OHO3N2. This figure is the same as \reffig{f.masssize}, except for
    the adopted metallicity calibration. In the parent sample we find that
    \OHO3N2 does not align with lines of constant \Pot (black dashed lines;
    panels~\subref{f.masssize.o3n2.a} and \subref{f.masssize.o3n2.f}). However,
    for each of the aperture-matched subsamples, the lines of constant \OHO3N2
    are always approximately aligned along lines of constant \Pot
    (panels~\subref{f.masssize.o3n2.b}~-~\subref{f.masssize.o3n2.e} and
    \subref{f.masssize.o3n2.g}~-~\subref{f.masssize.o3n2.j}). This behaviour is
    unique to the \OHO3N2 calibration, because the difference in the zero-point
    of the \PZR relations of each aperture-matched sample is highest for this 
    calibration.
    }\label{f.masssize.o3n2}
  \end{figure}

The only major difference is the behaviour of the \OHO3N2 distribution on the
mass-size plane: in the top row of \reffig{f.masssize.o3n2}
(panels~\subref{f.masssize.o3n2.a} and \subref{f.masssize.o3n2.f}), we find that
the lines of constant metallicity (lines of approximately uniform colour hue)
are not aligned with the lines of constant \Pot (dashed black lines). In contrast
to the results for the other calibrations, this observation is true both above and below the solid
white curve, marking the locus of average fibre coverage equal to 
$\mathrm{R_{fib}/R_e} = 0.5$. However, when we observe each of the aperture-matched
subsamples, we find that the lines of constant metallicity are aligned with the
lines of constant \Pot, as observed for all the other calibrations.
(panels~\subref{f.masssize.o3n2.b}~-~\subref{f.masssize.o3n2.e} and
\subref{f.masssize.o3n2.g}~-~\subref{f.masssize.o3n2.j} in Figs~\ref{f.masssize} and
\ref{f.masssize.o3n2}). The reason for this apparent discrepancy is that
the zero-point differences of the aperture-matched subsamples for the \PZR
relations are much greater for the \OHO3N2 metallicity than for any of the other
calibrations: as a result, mixing together different aperture-matched samples in
the parent sample introduces a stronger aperture-bias in
\reffig{f.masssize.o3n2.f} than in \reffig{f.masssize.f}. This emphasizes the
importance of aperture-matched sampling.

We conclude that our main results are independent of the metallicity calibration
adopted.


\section{Discussion}\label{disc}

In the previous section we have studied the correlation between the gas-phase
metallicity (\OH) and the three photometric estimators for mass \M,
gravitational potential \Pot and surface mass density \Sig. Here we attempt to
draw a consistent physical picture for the metallicity relations. We start
by comparing our results to other studies of the global and local metallicity
relations (\refsec{global.works} and \refsec{local.works}). We then compare the
gas-phase metallicity trends to what has been observed for the stellar
metallicity (\refsec{stellar.works}). Finally, we present two physical
interpretations in \refsec{phys.inter}.

\subsection{The aperture-averaged metallicity relations}\label{global.works}

In this work we improved upon previous studies in two critical aspects. Firstly,
by constructing a volume-limited sample, we ensure that our
results are as representative as possible of the galaxy population in the local
Universe (\refsec{ds.sprop}). Secondly, by studying the metallicity relations
within aperture-matched samples, we minimise aperture bias.

The importance of a volume-limited approach is
underlined by the controversy around the fundamental metallicity relation
\citetext{\citealp{mannucci+2010, sanchez+2013, salim+2014, telford+2016};~
\citetalias{sanchez+2017};~\citetalias{barrera-ballesteros+2018}}. 
A specific limitation of studying the \MZR relation in a volume-limited sample
is that our results are not directly comparable to the current literature, which
adopts different selection criteria \citetext{e.g. \citetalias{tremonti+2004};
\citealp{telford+2016}, \citetalias{sanchez+2017}}. For instance,
we find an RMS of $0.135$, larger than the values reported for both
fibre-based studies \citepalias[$0.10 \, \mathrm{dex}$; e.g.][]{tremonti+2004}
or IFU-based studies \citepalias[$0.06-0.10 \, \mathrm{dex}$; e.g.][]{sanchez+2017}.
However, this discrepancy is entirely explained by the characteristics of our
sample and by the metallicity calibration adopted.
Our volume-limited sample contains a larger fraction of low-mass galaxies
($10^9 < \mathrm{M_*} < 10^{10} \; \mathrm{M_\odot}$) compared to both
\citetalias{tremonti+2004} and \citetalias{sanchez+2017}: the \MZR relation
has larger RMS at the low-mass end than at the high-mass end
\citetext{\citet{guo+2016}, but see also \citet{kewley+ellison2008},
top panel of their fig.~2}. For this reason, the observed RMS depends primarily
on the fraction of low-mass galaxies.
For example, by adopting a volume-limited sample with
$z \leq 0.08$ and $\mathrm{M_*} > 10^{9.5} \, \mathrm{M_\odot}$, we find
$RMS=0.11$. When we repeat our analysis for the comparison samples, we find
results that are consistent with the published values (\refapp{app.a}).
As for the results obtained using IFU data \citepalias[e.g. ][]{sanchez+2017}, the
different mass function of our sample is compounded by the different quality of
the data: synthetic aperture spectra constructed from IFU spectroscopy typically
have much higher SNR than single-fibre spectroscopy.
In addition, the metallicity calibration of \citetalias{dopita+2016} has larger
scatter than other calibrations, as already demonstrated by other authors
\citepalias{sanchez+2017, barrera-ballesteros+2018}.
We conclude therefore that: (i) for any given metallicity calibration, the
average scatter about the \MZR relation is larger than the typical values
reported in the literature, which are based on magnitude-limited samples and
(ii) given that the increased RMS in the \MZR relation is due to sample
selection and metallicity calibration, it does not affect our comparative analysis
between the three metallicity relations considered here.
For this reason we judge that the
observed residual trend with size at fixed \M reflects an underlying physical
trend. Our conclusion is confirmed by other studies: the existence of a residual
correlation between \Z and galaxy size at fixed mass has already been pointed
out in previous works, both for the gas-phase metallicity \citep{ellison+2008,
salim+2014, telford+2016} as well as for the stellar metallicity
\citetext{\citealp{mcdermid+2015, scott+2017}; \citetalias{barone+2018};
\citealp{li+2018}}. Our results are in qualitative agreement with the
current literature, however we are the first to quantify the size-metallicity
relation for the gas.

The importance of the aperture-matched samples is highlighted by the
different behaviour of the metallicity distribution on the mass-size plane
above and below the line of fibre coverage $\mathrm{R_{fib}/R_e} = 0.5$.
In the region above this line, metallicity appears to follow mass, whereas
in the region below it metallicity follows potential. Given that this line also
marks the region below which we can trust our metallicity and \Pot measurements,
we argue that aperture-matched sampling enables us to overcome aperture bias
affecting previous fibre-based studies.

\subsection{The local metallicity relations}\label{local.works}

Alongside the aperture-averaged \MZR relation, several studies have found a correlation
between the local gas-phase metallicity \Zx and a number of local properties of
galaxies, typically defined on a physical scale of $1 \, \mathrm{kpc}$: local
stellar mass surface density \Sigstarx \citep{rosales-ortega+2012,
sanchez+2013, barrera-ballesteros+2016}, local escape velocity \vescx
\citepalias{barrera-ballesteros+2018} and local gas fraction
\fgasx$\equiv
\Sigma_\mathrm{gas}(\mathbf{x}) / (\Sigma_{gas}(\mathbf{x}) + \Sigma_*(\mathbf{x}))$,
with $\Sigma_\mathrm{gas}$ estimated from the Balmer decrement
\citepalias{barrera-ballesteros+2018}.
In addition, if one uses the average metallicity and the effective radius of
each galaxy as units of metallicity and radius, all galaxies have an average
radial gradient of $\alpha_\mathrm{O/H}=-(0.10 \pm 0.01) \; [\mathrm{dex/R_e}]$
\citep[radius-metallicity relation (\RZR);][]{sanchez+2014, ho+2015}.
Given that \fgasx, \Sigstarx and \vescx are correlated with one another, and
that they can all be expressed as declining exponentials in radius, it is
possible that one or more of the local relations are not independent.

Various authors have tried to explain which relations are physically motivated
\citetext{e.g. \citealp{barrera-ballesteros+2016};~\citetalias{barrera-ballesteros+2018}},
however we call into question the explanations offered to date.
Firstly, each of the three measurements comes with different random and
systematic uncertainties, therefore a direct comparison of the scatter about
the best-fit metallicity relations cannot be used to assess which relation is
intrinsically the tightest. We remark that even comparing the intrinsic scatter
would be controversial, because such a comparison relies on the accurate estimate
of the uncertainties.
At this stage, the best observational tool to find the most fundamental relation
is the study of residual trends. \citetalias{barrera-ballesteros+2018} suggest
that \fgasx is more important than \vescx because the residuals about the
best-fit \vescx-\Zx relation correlate strongly with \fgasx, whereas the
residuals about the \fgasx-\Zx relation have a relatively weak correlation with
\vescx. Although this argument is suggestive, there is no quantitative assessment
of the relative strength of the residual correlations. In addition,
\citepalias{barrera-ballesteros+2018} estimate \fgasx from the Balmer decrement
(see \refapp{app.c}),
which correlates observationally with metallicity via the dust fraction
\citep[][see also \refapp{app.c}]{draine+2014, groves+2015}.
With our fibre spectra we cannot address these issues directly. However
for stellar metallicity, the local \Mgbx-\vescx relation is mirrored by a
an aperture-averaged \Mgb-\vesc relation within one \re \citep{scott+2009}.
Arguably the same could be true for the gas metallicity.

\subsection{Comparison to stellar metallicity}\label{stellar.works}

The results shown here for the gas-phase metallicity are in qualitative
agreement with the trends of stellar metallicity (\ZH) for both
early-type and late-type galaxies \citep[ETGs, LTGs;][]{hubble1936, sandage1961}.
For ETGs, the relation between
\ZH and dynamical or stellar mass presents clear residual trends with galaxy
size \citep{mcdermid+2015, scott+2017} - in agreement with what we find here
for \OH for star-forming galaxies (\refsec{res.pzr}; \reffig{f.oh.a}).
\citetalias{barone+2018} have shown quantitatively that for ETGs, \ZH
correlates best with \Pot rather than with \M or \Sig, regardless of the
proxies (photometric or spectroscopic) used to estimate \M, \Pot and \Sig.
\ZH appears to follow lines of constant $\sigma$ (constant \Pot in our
terminology) also for LTGs \citep{li+2018}.

This qualitative agreemeent is particularly surprising for two reasons. Firstly,
the underlying physics between stellar and gas-phase metallicity are different:
stellar metallicity traces elements synthesised in all supernova types, whereas 
gas-phase metallicity (as measured here) traces only $\alpha$-elements, produced
only in type-II supernovae. Moreover, the timescale over which stellar and
gas-phase metallicity
evolve is also different: stellar metallicity measures the fraction of metals
locked in the atmosphere of all stars, therefore it reflects the integrated
star-formation history of a galaxy; gas-phase metallicity measures instead the
instantaneous abundance of metals in the star-forming gas, therefore it is
influenced by gas inflows and by recent star-formation
\citep[see e.g.][]{gonzalezdelgado+2014}. With these caveats in mind, a
qualitative comparison is still instructive.

A number of works have explored the dependence of local stellar metallicity on
local galaxy properties: \citet{emsellem+1996} find a relation between the local
value of the Lick index \Mgbx and the local escape velocity (\vescx).
\citet{scott+2009} observed that for the ETGs in the ATLAS$^\mathrm{3D}$ Survey,
the local and aperture-averaged relations between \Mgb and \vesc are consistent
across all galaxies.
If we interpret \Mgb as a crude proxy for \ZH, we expect a relation between the
local stellar metallicity and \vescx.

\citetalias{barone+2018} argue that the \ZH-\Pot relation indicates that most
stars in ETGs have formed \textit{in situ,} which in this context means in a
gravitational potential that is a
monotonic function of the current gravitational potential. For the stellar
metallicity, this relation can be maintained even through dry mergers
because the stars from accreted satellites are dispersed at a radius
where the local potential is equal to the binding energy of the satellite
\citep{villumsen1983, mcdermid+2015}.
In a closed-box model, the \textit{in situ} hypothesis predicts that the gas-phase
metallicity follows the same relation as the stellar metallicity, as indeed we
find in this study. Given that a closed-box model is not realistic, our results
suggest that either outflows are not as important in determining the gas-phase
metallicity, or they are regulated to a certain degree by the local
gravitational potential.

\subsection{Interpretation}\label{phys.inter}

The \MZR relation is widely understood to be shaped by two physical processes:
metal production and metal loss \citetext{e.g. \citetalias{tremonti+2004};
\citealp{lilly+2013, ho+2015}; \citetalias{barrera-ballesteros+2018}}.
To first order, we can assume that the mass
of metals produced is proportional to the stellar mass, therefore the gas-phase
metallicity \OH must be proportional to the ratio $\mathrm{O/H} \propto \mu^{-1}
\equiv \Sigma_*(\mathbf{x})/\Sigma_\mathrm{gas}(\mathbf{x})$; in this framework,
the local \Sigstarx-\OHx relation can
be interpreted as an enrichment sequence \citep[e.g.][]{barrera-ballesteros+2016}. By
integrating the \Sigstarx-\OHx relation (weighted with the light profile), one
finds the aperture-averaged \Sigstar-\OH relation. 
The second process, metal loss, depends on the value of \vesc \citepalias[e.g. ][]{
tremonti+2004}, which in turn is connected to the baryonic mass (estimated by
\Mstar) by
the baryonic Tully-Fisher relation \citep{bell+dejong2001}. The \MZR relation represents
therefore a sequence of metal retention. In practice \Mstar
and \Sigstar are strongly correlated, therefore both the \SZR and the \MZR
relations are jointly shaped by metal production and loss.
Given the evidence from these past works, the correlation between \OH and \Pot could
arise from the combination of two effects: (i) metal loss is related to \Mstar
\citepalias[as argued by ][thus explaining the \MZR relation]{tremonti+2004} and
(ii) metal enrichment is driven by \fgasx \citep[and therefore \Sigstarx, as argued by][thus
explaining both the local \Sigstarx-\OHx relation and the aperture-averaged \Sig-\OH
relation]{ho+2015, carton+2015, barrera-ballesteros+2016}.
Each of these effects induces a correlation between \OH and
$\mathrm{M_*/R_e^n}$, with $n=0$ and $n=2$ respectively. Their combination
produces an overall correlation with an intermediate value of $n$. In this
framework, the \RZR arises from the \Sigstarx-\OHx relation, and from the fact
that \Sigstar declines exponentially with radius for star-forming galaxies.
The presence of residual trends with galaxy size in both the
\MZR and the \SZR relations has a physical interpretation: at fixed \Sigstar,
larger galaxies have higher metallicity than smaller galaxies, because they are
more massive and hence have lower escape fraction. Conversely, at fixed \M,
larger galaxies (which have lower \Sigstar) have lower metallicity than smaller
galaxies (which have higher \Sigstar), because for a fixed escape fraction they
produced less metals per unit mass.

Although this explanation is attractive, it presents two difficulties.
Firstly, it requires some degree of fine tuning to explain why we find that the
best correlation with $\mathrm{M_*/R_e}^n$ has an exponent $n$ that is very
close to $1$, the value corresponding to the proxy for the gravitational
potential, \Pot.
Moreover, this hypothesis predicts that at the low-mass end of the relation
\Z correlates best with \M (because for low mass galaxies the metallicity is
dominated by outflows). Conversely, at the high-mass end, \Z should correlate
best with \Sigstar, because the escape fraction is small \citep[the escape fraction
is $f \lesssim 0.10$ for $\mathrm{M_* \gtrsim 3 \times 10^{10} \; M_\odot}$;][]{
ma+2016}. Neither of these predicted trends is observed in our sample: the lines
of constant gas-phase metallicity appear aligned along lines of constant \Pot
at every mass and size (\refsec{res.msp}).

Alternatively, the \PZR relation might point to a direct physical link between the average
depth of the gravitational potential \Pot and the average metallicity \Z. 
Firstly, if we consider \MstarRe as a crude proxy for the average \vesc, the
\PZR relation reflects the observed link between the local metallicity \Zx and the local
escape velocity \vescx. The existence of a link between \vescx and \Zx
\citepalias{barrera-ballesteros+2018} is in agreement with the observations for
the stellar \Mgb-\vesc relation \citep{emsellem+1996, scott+2009}.
In addition, if one accepts that the Balmer decrement is a good
estimator of \Siggas, we find that \Siggas is correlated more tightly with \Pot
than with either \Sigstar or \M (\refapp{app.c}).
Theoretically, the dynamical time of galaxies matches closely the timescale of
gas-consumption, \citep[e.g.][]{lilly+2013, torrey+2017}, which suggests that
the gravitational potential might play a key role in regulating star formation
and therefore metallicity.

\section{Summary and Conclusions}\label{conc}

In this work we examined the dependence of gas-phase metallicity (\OH) on the
physical size of galaxies (\re) for a sample of $\sim \, 70000$ local star-forming
galaxies drawn from the SDSS DR7. We use both a volume-limited parent sample,
comprising galaxies with different radial coverage and three
aperture-matched subsamples, selected to have a homogeneous fractional area
coverage from the fixed-aperture SDSS fibres. For each sample, we compare \OH to
three structural parameters: mass \M (estimated by \Mstar), average gravitational
potential \Pot (estimated by \MstarRe) and surface mass density \Sig (estimated
by \MstarReT). Our results are as follows:

\begin{itemize}
  \item We demonstrate the use of aperture-matched sampling based on single-fibre
    spectroscopy to study radially-varying properties of galaxies, while minimising
    aperture bias.
  \item We show the existence of a size-metallicity relation at fixed
    mass, independent of aperture bias (\refsec{res.fix};
    \reffig{f.fixed}).
    The logarithmic slope of the size-metallicity relation is opposite in sign,
    and has comparable absolute value to, the logarithmic slope of the
    mass-metallicity relation (\refsec{res.fix}; \reftab{t.fixed}).
  \item Over the region of the mass-size plane where we can reliabily measure the
    metallicity, the lines of constant metallicity are closely aligned
    with the lines of constant \Pot (\refsec{res.msp}; \reffig{f.masssize}).
  \item We find that - regardless of the sample selection adopted - the \M-\OH
    and \Pot-\OH relations have comparable root mean square (RMS) and Spearman rank
    correlation coefficient (\rhosp; \reffig{f.oh.a}-\subref{f.oh.b}).
  \item For each of the aperture-matched subsamples, the \Pot-\OH relation has the
    smallest RMS, the highest \rhosp and the least-significant residual
    trends with galaxy size (\reffig{f.oh.d}-\subref{f.oh.l}).
  \item We explore two possible explanations for the dependence of \OH on the
    physical parameters \M, \Pot and \Sig. The usual theory offered in the
    literature explains the \Sig-\OH and \M-\OH relations in terms of local
    enrichment (driven by \Sigstarx) and global escape fractions (driven by
    \M). We suggest, as an alternative, that both the \M-\OH and \Sig-\OH
    relations arise from a local relation between \OH and \Pot. This hypothesis
    can be tested using integral field spectroscopy.
\end{itemize}


\section*{Acknowledgements}

FDE acknowledges useful discussion with Prof. Roger L. Davies, Dr. Luca Cortese,
Prof. Andrea Macci\`o, Dr. Chiaki Kobayashi, Dr. Philip Taylor,
Dilyar Barat and with the members of the SAMI Galaxy
Survey team. TMB is supported by an Australian Government Research Training
Program Scholarship.
Parts of this research were supported by the Australian Research Council
Centre of Excellence for All-sky Astrophysics (CAASTRO; grant CE110001020),
and the Australian Research Council Centre of Excellence for All Sky Astrophysics
in 3 Dimensions (ASTRO 3D; grant CE170100013).
BG gratefully acknowledges the support of the Australian Research Council as the
recipient of a Future Fellowship (FT140101202).

Funding for the SDSS and SDSS-II has been provided by the Alfred P. Sloan
Foundation, the Participating Institutions, the National Science Foundation, the
U.S. Department of Energy, the National Aeronautics and Space Administration,
the Japanese Monbukagakusho, the Max Planck Society, and the Higher Education
Funding Council for England. The SDSS Web Site is http://www.sdss.org/.\\

The SDSS is managed by the Astrophysical Research Consortium for the
Participating Institutions. The Participating Institutions are the American
Museum of Natural History, Astrophysical Institute Potsdam, University of Basel,
University of Cambridge, Case Western Reserve University, University of Chicago,
Drexel University, Fermilab, the Institute for Advanced Study, the Japan
Participation Group, Johns Hopkins University, the Joint Institute for Nuclear
Astrophysics, the Kavli Institute for Particle Astrophysics and Cosmology, the
Korean Scientist Group, the Chinese Academy of Sciences (LAMOST), Los Alamos
National Laboratory, the Max-Planck-Institute for Astronomy (MPIA), the
Max-Planck-Institute for Astrophysics (MPA), New Mexico State University, Ohio
State University, University of Pittsburgh, University of Portsmouth, Princeton
University, the United States Naval Observatory, and the University of
Washington.\\

This work made extensive use of the Debian GNU/Linux operative system, freely
available at http://www.debian.org. We used the Python programming language
\citep{vanrossum1995}, maintained and distributed by the Python Software
Foundation, and freely available at http://www.python.org. We acknowledge the 
use of scipy \citep{jones+2001}, matplotlib \citep{hunter2007}, emcee
\citep{foreman-mackey+2013}, astropy \citep{astropyco+2013} and pathos
\citep{mckerns+2011}.
During the preliminary analysis we have made extensive use of TOPCAT
\citep{taylor2005}.

\selectlanguage{english}


\bibliographystyle{mnras}
\bibliography{astrobib}


\appendix

\section{Notable samples}\label{app.a}

We favour a volume-limited sample to ensure that our conclusions are
representative of the local Universe (\refsec{ds}). However the results shown
in \refsec{res} are largely independent of the sample selection criteria. Here
we repeat our analysis for two alternative samples selected following the
criteria of two popular studies of the mass-metallicity relation:
\citetalias{tremonti+2004} and of \citet{telford+2016}.

  \begin{figure*}
    \centering
    \includegraphics[type=pdf,ext=.pdf,read=.pdf,width=1.0\textwidth]{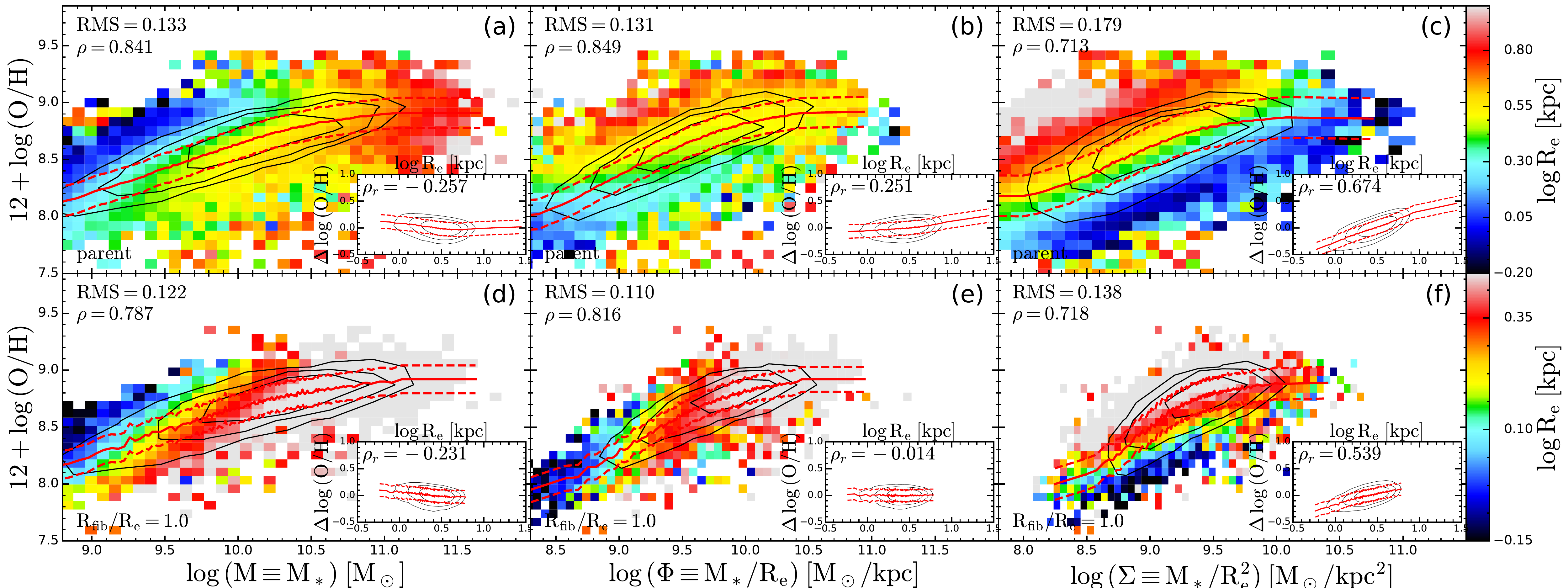}
    {\phantomsubcaption\label{f.oh.Tre04.a}
     \phantomsubcaption\label{f.oh.Tre04.b}
     \phantomsubcaption\label{f.oh.Tre04.c}
     \phantomsubcaption\label{f.oh.Tre04.d}
     \phantomsubcaption\label{f.oh.Tre04.e}
     \phantomsubcaption\label{f.oh.Tre04.f}}
    \caption{The same as \reffig{f.oh}, but for a sample reproducing the
    selection of \citetalias{tremonti+2004}. The top row shows gas-phase metallicity
    as a function of \M, \Pot and \Sig for the parent sample (panels~
    \subref{f.oh.Tre04.a}~-~\subref{f.oh.Tre04.c}). The bottom row shows the
    results for the aperture-matched subsample with $\mathrm{R_{fib}/R_e} = 1$
    (\subref{f.oh.Tre04.d}~-~\subref{f.oh.Tre04.f}). For the parent sample, the
    best predictor of \OH is \M (it has less scatter, higher \rhosp and less
    residual trends with size than the \OH-\Pot and \OH-\Sig relations). For the
    aperture-matched subsample, the best predictor of \OH is \Pot.
    }\label{f.oh.Tre04}
  \end{figure*}

\citetalias{tremonti+2004} selected their sample imposing: (i) $0.005 < z < 0.25$,
(ii) $\mathrm{SNR(H\alpha)} \geq 5$, $\mathrm{SNR(H\beta)} \geq 5$ and
$\mathrm{SNR(NII\lambda6584)} \geq 5$, (iii) uncertainty on $z-$band magnitude
$\sigma(m_z) < 0.15 \; \mathrm{mag}$, (iv) uncertainty on the $\mathrm{H}\delta_A$
index $\sigma(\mathrm{H}\delta_A) < \SI{2.5}{\angstrom}$, (v) uncertainty on the
$D_n(4000)$ index $\sigma(D_n(4000)) < 0.1$, (vi) classified as star-forming
according to the BPT diagram of \citet{kauffmann+2003c} and (vii) uncertainty on
stellar mass and metallicity less than $0.2 \; \mathrm{dex}$. Here we reproduce
their selection, but with two differences: we use data from SDSS DR7 instead of
SDSS DR4, and we use the metallicity calibration from \citet[][see \refsec{ds.metal}]{dopita+2016}. Despite the
different selection criteria between this sample and our sample (\refsec{ds}),
the results of the comparative analysis are the same. In \reffig{f.oh.Tre04}
we compare \OH to \M, \Pot and \Sig for the full sample (top row,
panels~\subref{f.oh.Tre04.a}~-~\subref{f.oh.Tre04.c}) and for the
aperture-matched subsample with $\mathrm{R_{fib}/R_e} = 1$ (bottom row,
panels~\subref{f.oh.Tre04.d}~-~\subref{f.oh.Tre04.f}). For the full sample,
the \M-\OH and \Pot-\OH relations are equivalent: they have comparable RMS and \rhosp
and the residuals have similar but opposite correlations with galaxy size.
However, for the aperture-matched samples, the best predictor of \OH is \Pot, as
we have found in \refsec{res} (Figs~\ref{f.oh.d}~-~\subref{f.oh.f}).

  \begin{figure*}
    \centering
    \includegraphics[type=pdf,ext=.pdf,read=.pdf,width=1.0\textwidth]{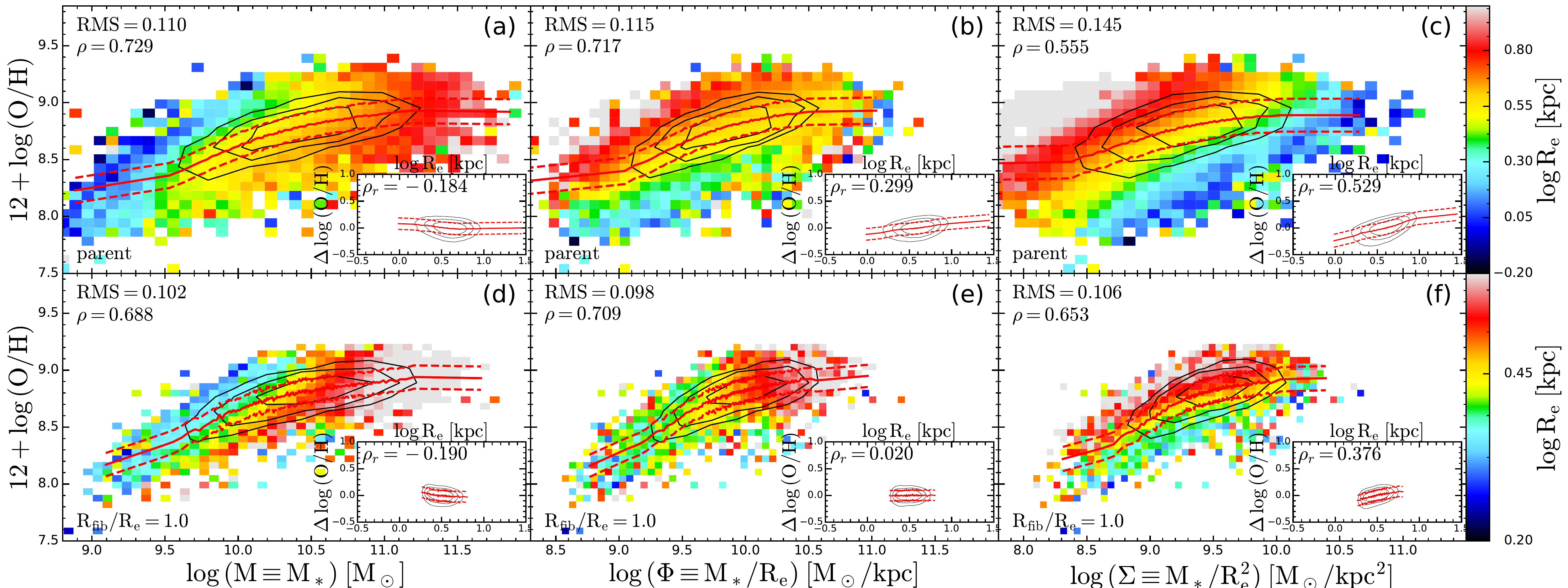}
    {\phantomsubcaption\label{f.oh.Tel16.a}
     \phantomsubcaption\label{f.oh.Tel16.b}
     \phantomsubcaption\label{f.oh.Tel16.c}
     \phantomsubcaption\label{f.oh.Tel16.d}
     \phantomsubcaption\label{f.oh.Tel16.e}
     \phantomsubcaption\label{f.oh.Tel16.f}}
    \caption{The same as \reffig{f.oh}, but for a sample reproducing the
    selection of \citet{telford+2016}. The top row shows gas-phase metallicity
    as a function of \M, \Pot and \Sig for the parent sample (panels~
    \subref{f.oh.Tel16.a}~-~\subref{f.oh.Tel16.c}). The bottom row shows the
    results for the aperture-matched subsample with $\mathrm{R_{fib}/R_e} = 1$
    (\subref{f.oh.Tel16.d}~-~\subref{f.oh.Tel16.f}). For the parent sample, the
    best predictor of \OH is \M (it has less scatter, higher \rhosp and less
    residual trends with size than the \OH-\Pot and \OH-\Sig relations). For the
    aperture-matched subsample, the best predictor of \OH is \Pot.
    }\label{f.oh.Tel16}
  \end{figure*}

\citet{telford+2016} selected their sample to minimize the bias against low
metallicity galaxies. Their constraints are as follows: (i) $0.07 < z < 0.30$,
$A_V < 2.5$, (ii) Balmer decrement $F(\mathrm{H\alpha})/F(\mathrm{H\beta}) >
2.5$, (iii) $\mathrm{SNR(H\alpha)} \geq 25$, $\mathrm{SNR(H\beta)} \geq 5$,
$\mathrm{SNR(SII\lambda6717)} \geq 3$ and $\mathrm{SNR(SII\lambda6731)} \geq 3$,
(iv) classified as star-forming according to the BPT diagram of
\citet{kauffmann+2003c} and (v) having valid size measurements in the catalogue of
\citet{simard+2011}. These selection criteria are identical to ours
(\ref{ds.samplesel}), except for the redshift range: the volume considered by
\citet{telford+2016} is $\approx 80$ times larger than ours, but the low mass
range is incomplete and would require large corrections, which we chose to avoid
in this work.

Despite the different mass and redshift range, and without applying any
correction for incompleteness, using this sample leads to the same conclusions
as our volume-limited approach: the outcome of the comparative analysis
between the metallicity relations with \M, \Pot and \Sig is the same as for our
sample (\reffig{f.oh.Tel16}). For the full sample, the \M-\OH relation appears
to be the best predictor of metallicity: it has lower RMS and higher \rhosp
than both the \Pot-\OH and \Sig-\OH relations
(panels~\subref{f.oh.Tel16.a}~-~\subref{f.oh.Tel16.c}). Even though the
residuals of the \M-\OH relation are correlated with \logre ($\rho_r = -0.184$;
inset panel in \reffig{f.oh.Tel16.a}), this residual correlation is weaker than
the residual correlations for the \Pot-\OH and \Sig-\OH relations (inset panels
in Figs~\ref{f.oh.Tel16.b} and \subref{f.oh.Tel16.c}).

However, when we study
the relation for the aperture-matched subsample with $\mathrm{R_{fib}/R_e} = 1$,
we find that it is the \Pot-\OH relation which is the best predictor of \OH: it
has the lowest RMS and the highest \rhosp. Moreover, the residuals of the
\Pot-\OH relation show no correlation with \logre ($\rho = 0.020$; inset panel
in \reffig{f.oh.Tel16.e}), unlike the residuals of the \M-\OH and \Sig-\OH
relations ($\rho = -0.190$; inset panel in \reffig{f.oh.Tel16.d} and
$\rho = 0.376$; inset panel in \reffig{f.oh.Tel16.f}).

This analysis reinforces our conclusion that our results do not depend on the
sample selection criteria and that aperture-matched sampling is critical to
recover an unbiased relation (i.e. aperture bias is more important than how
the mass function is sampled).

\section{Systematic variations in the light profile}\label{app.b}

One major concern of our analysis is that studies stemming from IFU spectroscopy
recommend using aperture-averaged metallicities within $2 \; \mathrm{R_e}$
\citep{iglesias-paramo+2016}, but here the corresponding aperture-matched sample
(i.e. $\mathrm{R_{fib}/R_e} = 2$) does not possess enough galaxies to perform a
conclusive quantitative analysis.
\citet{iglesias-paramo+2016} argue that the systematic variation in the light
profile among different spiral types introduces a morphology-dependent aperture
bias. Early spirals (Sa) have by definition more prominent bulges than late
spirals \citep[Sc, Sd;][]{hubble1936, sandage1961}.
As a result, the former have
more concentrated light profiles than the latter, which in turn means that a
smaller fraction of the disc light is included within $1 \, \mathrm{R_e}$ than
for later-type galaxies.

In order to assess whether this bias is affecting our results, we used three
methods. Firstly, we have shown qualitatively that even for the aperture-matched
sample with $\mathrm{R_{fib}/R_e} = 2$, \OH closely follows lines of
constant \Pot (\reffig{f.masssize.j}). Secondly, we show that our quantitative
results hold for the sample with $\mathrm{R_{fib}/R_e} = 1.5$
(panels~\subref{f.oh.j}-\subref{f.oh.l} in \reffig{f.oh}). Here we show that
even by selecting galaxies with the same light profiles, our results are unchanged.

  \begin{figure*}
    \centering
    \includegraphics[type=pdf,ext=.pdf,read=.pdf,width=1.0\textwidth]{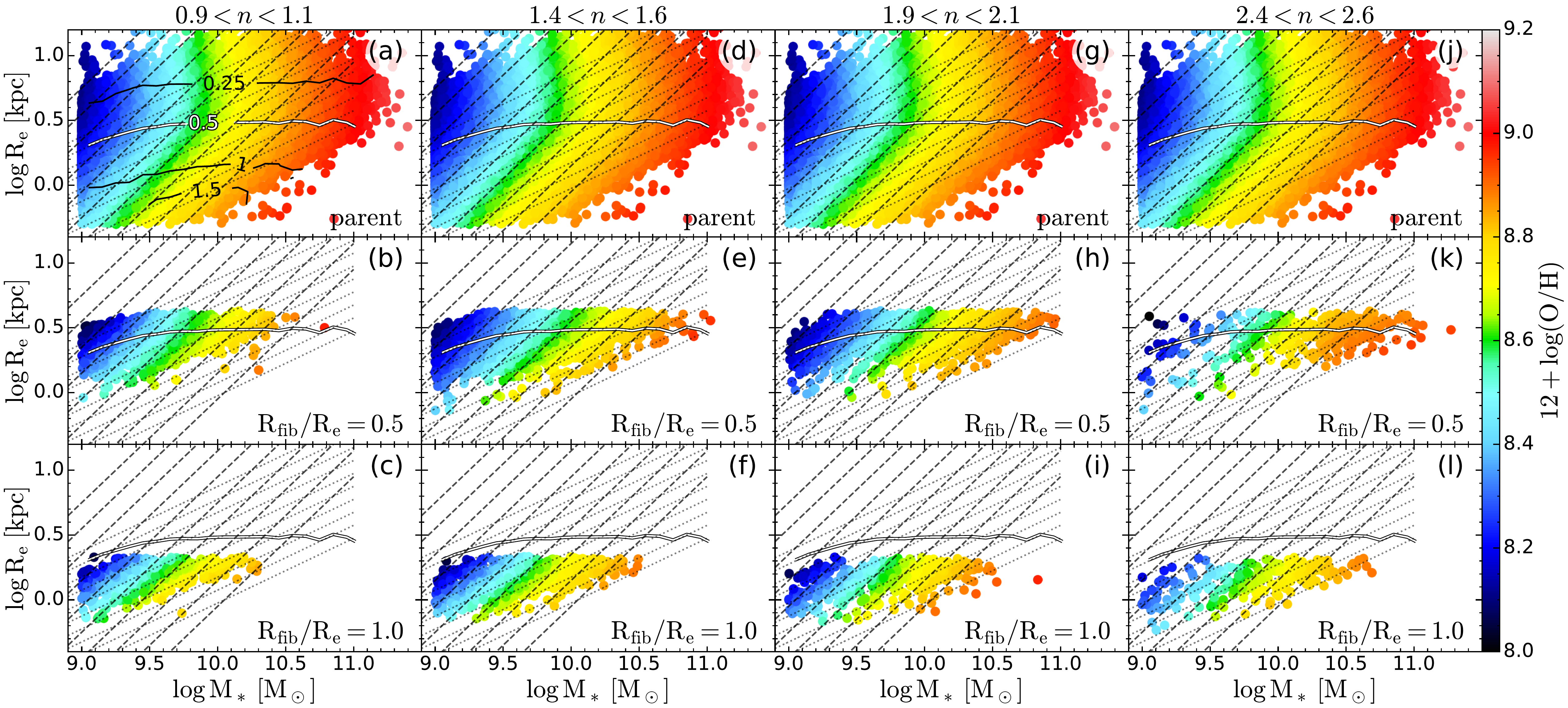}
    {\phantomsubcaption\label{f.masssize.sersic.a}
     \phantomsubcaption\label{f.masssize.sersic.b}
     \phantomsubcaption\label{f.masssize.sersic.c}
     \phantomsubcaption\label{f.masssize.sersic.d}
     \phantomsubcaption\label{f.masssize.sersic.e}
     \phantomsubcaption\label{f.masssize.sersic.f}
     \phantomsubcaption\label{f.masssize.sersic.g}
     \phantomsubcaption\label{f.masssize.sersic.h}
     \phantomsubcaption\label{f.masssize.sersic.i}
     \phantomsubcaption\label{f.masssize.sersic.j}
     \phantomsubcaption\label{f.masssize.sersic.k}
     \phantomsubcaption\label{f.masssize.sersic.l}}
    \caption{Distribution of the gas-phase metallicity \OH on the mass-size
    plane, divided by the best-fit value of the $r-$band S{\'e}rsic index $n$.
    Each column shows the distribution for a different bin in $n$: $0.9 < n <
    1.1$ (panels~\subref{f.masssize.sersic.a}-\subref{f.masssize.sersic.c}),
    $1.4 < n < 1.6$ (panels~\subref{f.masssize.sersic.d}-\subref{f.masssize.sersic.f}),
    $1.9 < n < 2.6$ (panels~\subref{f.masssize.sersic.g}-\subref{f.masssize.sersic.i}),
    $2.4 < n < 2.6$ (panels~\subref{f.masssize.sersic.j}-\subref{f.masssize.sersic.l}).
    Each row shows the distribution for a different aperture: the parent sample
    (top row, panels~\subref{f.masssize.sersic.a}, \subref{f.masssize.sersic.d},
    \subref{f.masssize.sersic.g} and \subref{f.masssize.sersic.j}), the
    aperture-matched sample with $\mathrm{R_{fib}/R_e} = 0.5$ (middle row,
    panels~\subref{f.masssize.sersic.b}, \subref{f.masssize.sersic.e},
    \subref{f.masssize.sersic.h} and \subref{f.masssize.sersic.k}) and the
    aperture-matched sample with $\mathrm{R_{fib}/R_e} = 1$ (bottom row,
    panels~\subref{f.masssize.sersic.c}, \subref{f.masssize.sersic.f},
    \subref{f.masssize.sersic.i} and \subref{f.masssize.sersic.l}). In the
    aperture-matched samples, \OH is constant (i.e. approximately uniform
    colour hue) along lines of constant \Pot (dashed black lines). This fact
    does not depend on the functional shape of the stellar light profile.
    }\label{f.masssize.sersic}
  \end{figure*}

If the \PZR relation were due to light-profile dependent aperture bias still
being present in our aperture-matched samples, we would expect that dividing
these samples by the shape of the light profile showed a different dependence
on mass and size. In \reffig{f.masssize.sersic} we reproduce the metallicity
variation on the mass-size plane reconstructed with LOESS, divided in
four narrow bins of the S{\'e}rsic index $n$: $0.9 < n < 1.1$ (first column),
$1.4 < n < 1.6$ (second column), $1.9 < n < 2.1$ (third column) and
$2.4 < n < 2.6$ (last column; for higher values of $n$ our aperture-matched
samples do not possess sufficient galaxies). Comparing
\reffig{f.masssize.sersic} with \reffig{f.masssize} we find no qualitative
difference: galaxies with different light profiles behave in the same way on
the mass-size plane, so the fact that the lines of constant \OH are
parallel to the lines of constant \Pot is not an artefact of aperture bias.

\section{Using the Balmer decrement to estimate the gas surface mass density}\label{app.c}

  \citetalias{barrera-ballesteros+2018} use the Balmer decrement to estimate the gas surface mass
  density, by defining $\Sigma_\mathrm{gas} \equiv 30 \, A_V \;
  \mathrm{M_\odot \; pc^{-2}}$, where $A_V$ is the optical extinction. This
  estimator assumes that the dust-to-gas ratio ($\mathcal{D}$) is uniform across
  all galaxies, however $\mathcal{D}$ depends on metallicity \citep{draine+2014,
  groves+2015}.
  For this reason one should be careful when using this method to study the
  correlation between metallicity and the surface mass density of gas.

  \begin{figure}
    \centering
    \includegraphics[type=pdf,ext=.pdf,read=.pdf,width=1.0\columnwidth]{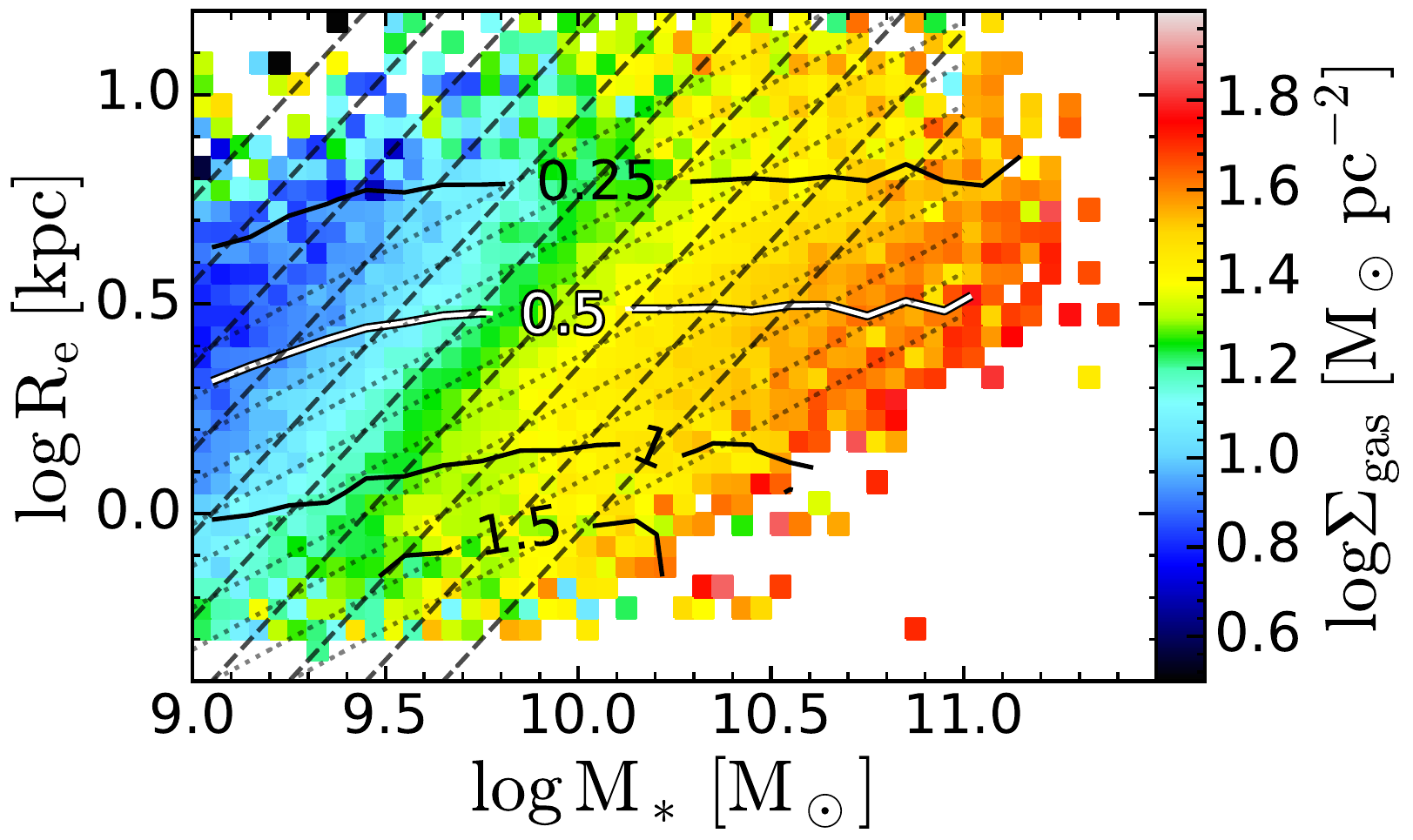}
    \caption{Gas surface density (\Siggas, derived from the Balmer decrement) on
    the mass size plane. The solid lines are lines of constant (median) fibre
    coverage $\kappa = \mathrm{R_{fib}/R_e}$. Our fibre-based measurements of
    metallicity can be trusted only in the region below the solid white curve
    ($\kappa = 0.5$). The lines of constant \Siggas (i.e. lines of uniform
    colour) are aligned to the lines of constant potential \Pot (dashed black
    lines) rather than the lines of constant surface density \Sig (dotted black
    lines).}\label{f.balmdec}
  \end{figure}

  In \reffig{f.balmdec} we show how $\Sigma_\mathrm{gas}$ (that is, $A_V$) varies
  across the mass-size plane. The solid lines are the loci of constant fibre
  coverage: our measurements of metallicity can be trusted only below the solid
  white curve (i.e. for $\mathrm{R_{fib}/R_e} > 0.5$). The dashed (dotted) black
  lines are lines of constant \Pot (\Sig): it is clear that \Siggas correlates
  with \Pot, rather than with \Sig. When we repeat the quantitative analysis of
  \reffig{f.oh} by substituting \Siggas in place of \OH,
  we find that the \Siggas-\Pot and \Siggas-\Sig relations have comparable RMS,
  but the \Siggas-\Pot relation is statistically consistent with no residual trends with
  galaxy size, unlike the \Siggas-\Sig relation.
  We believe that this result is a consequence of the correlation between \Pot
  and \OH, and between \OH and the Balmer decrement. If however we were to treat
  the Balmer decrement as an unbiased estimator of \Siggas, the fact that \Pot is
  the best predictor of \Siggas could be used to explain the \Pot-\OH relation in
  terms of effective yield and gas fractions
  \citepalias[e.g.][]{barrera-ballesteros+2018}.
  
  The fact that \Siggas correlates better with \Pot than with \M or \Sig raises
  the possibility that the \Pot-\OH relation arises from the correlation between
  \Pot and $A_V$ and from $A_V$ biasing the metallicity estimator \OH. However
  the metallicity calibration adopted here uses line ratios that are sufficiently
  close in wavelength to be insensitive to the extinction correction
  \citep[see \refsec{ds.metal} and ][]{dopita+2016}. For this reason we believe
  that \OH links \Pot and $A_V$, rather than $A_V$ linking \Pot and \OH.

\label{lastpage}

\end{document}